\documentclass[preprint,12pt,authoryear]{elsarticle}

\usepackage{amssymb}
\usepackage{amsmath}

\usepackage{latexsym}
\usepackage{bm}
\usepackage{booktabs}
\usepackage{makecell}
\usepackage{mathtools}
\usepackage{url}
\usepackage{xcolor}
\usepackage{hyperref}
\usepackage{graphicx}
\usepackage{float}
\usepackage{tabularx}
\usepackage{pgfplots}
\usepackage{caption}
\usepackage{threeparttable}
\usepackage{etoolbox}
\usepackage[utf8]{inputenc}
\usepackage{multirow}
\usepackage{pifont} 
\usepackage{placeins}

\newcommand{\cmark}{\ding{51}} 
\newcommand{\xmark}{\ding{55}} 
\AtBeginEnvironment{table}{\small} 

\begin{document}

\begin{frontmatter}



\title{Read Like a Radiologist: Efficient Vision-Language Model for 3D Medical Imaging Interpretation
} 



\author[gsaikaist]{Changsun Lee\fnref{fn1}} 
\author[severance]{Sangjoon Park\fnref{fn1}}
\author[snuh]{Cheong-Il Shin}
\author[stvinh]{Woo Hee Choi}
\author[cauh]{Hyun Jeong Park}
\author[cnuh]{Jeong Eun Lee\corref{cor1}}
\ead{leeje290@gmail.com}
\author[gsaikaist]{Jong Chul Ye\corref{cor1}}
\ead{jong.ye@kaist.ac.kr}

\affiliation[gsaikaist]{organization={Kim Jaechul Graduate School of AI, Korea Advanced Institute of Science and Technology (KAIST)},
            city={Daejeon},
            country={Republic of Korea}}
\affiliation[severance]{organization={Department of Radiation Oncology, Yonsei Cancer Center, Heavy Ion Therapy Research Institute, Yonsei University College of Medicine},
            city={Seoul},
            country={Republic of Korea}}
\affiliation[cnuh]{organization={Department of Radiology, Chungnam National University Hospital, Chungnam National University College of Medicine},
            city={Daejeon},
            country={Republic of Korea}}
\affiliation[snuh]{organization={Department of Radiology, Seoul National University Hospital, Seoul National University College of Medicine},
            city={Seoul},
            country={Republic of Korea}}
\affiliation[stvinh]{organization={Department of Nuclear Medicine, St. Vincent's Hospital, College of Medicine, The Catholic University of Korea},
            city={Suwon},
            country={Republic of Korea}}
\affiliation[cauh]{organization={Department of Radiology, Chung-Ang University Hospital, Chung-Ang University College of Medicine},
            city={Seoul},
            country={Republic of Korea}}
\cortext[cor1]{Corresponding authors}
\fntext[fn1]{Co-first author}








\begin{abstract} 

Recent medical vision-language models (VLMs) have shown promise in 2D medical image interpretation. However extending them to 3D medical imaging has been challenging due to computational complexities and data scarcity. Although a few recent VLMs specified for 3D medical imaging have emerged, all are limited to learning volumetric representation of a 3D medical image as a set of sub-volumetric features. Such process introduces overly correlated representations along the z-axis that neglect slice-specific clinical details, particularly for 3D medical images where adjacent slices have low redundancy. To address this limitation, we introduce MS-VLM that mimic radiologists' workflow in 3D medical image interpretation. Specifically, radiologists analyze 3D medical images by examining individual slices sequentially and synthesizing information across slices and views. Likewise, MS-VLM leverages self-supervised 2D transformer encoders to learn a volumetric representation that capture inter-slice dependencies from a sequence of slice-specific features. Unbound by sub-volumetric patchification, MS-VLM is capable of obtaining useful volumetric representations from 3D medical images with any slice length and from multiple images acquired from different planes and phases. We evaluate MS-VLM on publicly available chest CT dataset CT-RATE and in-house rectal MRI dataset. In both scenarios, MS-VLM surpasses existing methods in radiology report generation, producing more coherent and clinically relevant reports. These findings highlight the potential of MS-VLM to advance 3D medical image interpretation and improve the robustness of medical VLMs. 

\end{abstract}


\end{frontmatter}

\section{Introduction}

Attempts to non-invasively visualize the internal structures of the human body have driven significant advancements in medical imaging technology. These efforts began with two-dimensional (2D) modalities like plain X-rays, but gradually evolved toward more sophisticated three-dimensional (3D) imaging techniques, such as computed tomography (CT) and magnetic resonance imaging (MRI), which offer a more detailed visualization of anatomical structures~\citep{bushberg2011essential, kalender2011computed, mcrobbie2017mri}. CT and MRI imaging, by providing comprehensive insights without the need for invasive procedures, have become crucial and indeed inevitable tools in modern medicine, supporting both diagnosis and treatment across diverse medical fields~\citep{bushberg2011essential, kalender2011computed, mcrobbie2017mri}.

However, interpreting 3D medical imaging modalities, such as CT and MRI demands high level of expertise, often requiring significant time and effort from well-trained radiologists. Radiologists, however are not immune to the risk of error~\citep{brady2017error}, which has spurred the development of artificial intelligence (AI) models aimed at assisting radiologists' 3D medical image interpretation~\citep{litjens2017survey}. Despite some successes, particularly in specific tasks such as segmentation and detection~\citep{wang2019volumetric, hatamizadeh2022unetr, litjens2017survey}, when compared to the performance of multi-modal models on simpler 2D modalities like X-rays~\citep{lee2024llmcxr, li2024llava, chaves2024training}, these AI models still face significant limitations in broader applications such as radiology report generation or visual question answering (VQA).

The limited success of 3D imaging AI models can be attributed to several key factors. First, there is a fundamental shortage of paired 3D image-report datasets compared to the abundance of datasets available for 2D imaging modalities such as X-rays~\citep{mimiccxr, openi, irvin2019chexpert}, resulting in significant data insufficiency. This challenge is further exacerbated by the immense data volume and computational demands of 3D imaging, which are several hundred times greater than those of 2D imaging~\citep{litjens2017survey}. Second, the inherent complexity of 3D medical imaging poses significant challenges for effective encoding, as current 3D vision encoders often lack the capacity to capture the full scope of information required for accurate 3D image interpretation~\citep{litjens2017survey, dou20173d, isensee2021nnu}. Existing vision-language models (VLMs) also struggle to efficiently process the complex and voluminous data inherent in 3D imaging, leading to notable inefficiencies. Specifically, most state-of-the-art models~\citep{wu2023towards, blankemeier2024merlin, bai2024m3d} rely on 3D vision encoders that partition image volumes into fixed-size 3D patches. This approach introduces processing inefficiencies by creating unnecessary correlations across the z-axis, further limiting their effectiveness in handling 3D medical imaging data.

To address these challenges, we propose a novel VLM inspired by the way radiologists interpret 3D medical images. Unlike traditional methods that treat the entire 3D volume as a uniform entity, our model, termed MultiSlice Vision Language Model (MS-VLM), processes 3D volumes as a collection of 2D planes. This approach enables the model to encode information from each 2D plane independently before combining them to produce a comprehensive interpretation. By leveraging this method, MS-VLM achieves significantly higher data and computational efficiency while delivering superior performance compared to conventional approaches. Additionally, the structural design of MS-VLM allows it to handle volumetric data with variable slice lengths, enhancing its versatility and adaptability across diverse clinical applications. Notably, MS-VLM excels in scenarios where specific planes and phases are critical for accurate diagnosis, such as rectal MRI, where interpretations across different phases (e.g., pre-contrast and post-contrast) and planes (e.g., sagittal and axial) are essential. By incorporating reconstructions of images from multiple planes and phases, MS-VLM closely mirrors the interpretive processes employed by radiologists, leading to improved diagnostic performance.

Our contributions can be summarized as follows:

\begin{itemize}
   \item Inspired by radiologists, we propose a novel 3D imaging interpretation method that treats 3D volumes as a collection of 2D planes, significantly improving data and computational efficiency.
   
   \item  By introducing a novel Z-former, the proposed MS-VLM model handles volumetric data with variable slice lengths, enhancing its practical utility in diverse clinical scenarios.
   
   \item Specifically, by incorporating reconstructions from various planes and phases through the Z-former, MS-VLM aligns with radiologists' interpretative workflows, resulting in improved performance.
\end{itemize}

\section{Related Work}
In this section, we provide a concise overview of recent research on the development of multi-modal foundation models in the medical domain and the approaches of the VLM for 3D medical imaging.

\subsection{Multi-modal Foundation Model in Medical Domain}
Healthcare is inherently multi-modal, requiring clinicians to synthesize information from various sources, such as medical images, clinical notes, lab results, electronic health records (EHR), genomic data, etc. to provide effective patient care. In recent years, significant progress has been made in developing AI systems that achieve expert level performance within individual modalities, such as interpreting radiographic images, analyzing pathology slides, or detecting genetic mutations~\citep{pham2022accurate, taori2023stanford, qu2021genetic}. However, the complexity and multi-modal nature of healthcare require AI models capable of integrating these diverse data types into a cohesive framework. 
The early attempts to achieve this integration leverage contrastive learning models, such as CLIP \citep{radford2021learning}. CLIP maps images and corresponding contextual information into a shared embedding space through large-scale contrastive learning, demonstrating promising results in integrating simple medical imaging data, like chest radiographs, with radiology reports. Medical adaptions of CLIP such as MedCLIP, CheXzero and CXR-CLIP~\citep{wang2022medclip, tiu2022expert, you2023cxr} perform image-text retrieval on chest X-ray (CXR) images and the corresponding reports demonstrating bidirectional understanding of the two modalities. 

Building on this foundation, more sophisticated approaches emerged to directly integrate data from different modalities. Unlike earlier CLIP-based methods, these newer approaches enable a more comprehensive understanding of multi-modal data, supporting advanced applications such as detailed medical report generation and VQA. Vision encoders for image data and text encoders for textual information are used together, with cross-attention and self-attention mechanisms facilitating seamless integration of complex information. This deeper level of integration not only correlates visual and textual data but also supports holistic tasks that require a nuanced understanding of medical data. Notable examples include PPKED, MedViLL, and Medical X-VL\citep{liu2021exploring, moon2022multi, park2024self}. 

More recently, the impressive capabilities of large language models (LLMs) have driven efforts to develop multi-modal models with a generalist architecture. These models aim to handle multiple modalities using a single framework, enhancing flexibility and simplifying the integration process compared to specialized encoder modules. Med-PaLM M~\citep{tu2024towards} is a pioneering effort in this area, demonstrating the ability to process text, imaging, and genomic data within a unified framework. This design maximizes flexibility and information transfer across modalities, providing a strong foundation for future multi-modal AI systems. Med-PaLM M has also shown potential to outperform specialized models in specific tasks, suggesting that generalist models may offer superior performance even in domain-specific applications. Subsequent models, such as BioMedGPT and PathChat~\citep{zhang2024generalist, lu2024multimodal}, have further built on these capabilities, underscoring the potential of foundational models to provide generalized solutions in the medical domain. 

\subsection{VLMs for 3D Medical Imaging Interpretation}
Most initial efforts to develop effective multi-modal models in the medical domain have largely focused on simpler 2D imaging modalities, such as plain radiographs and pathology slides~\citep{tu2024towards, li2024llava}. 
However, developing such models for 3D medical imaging modalities, like CT and MRI, have presented unique challenges: limited training data and high data dimensionality. Furthermore, current vision encoders often lack the capacity to adequately process the intricate information found in 3D medical images, as they require handling large-scale data while understanding complex spatial relationships.

Unlike natural 3D data such as video, which contains significant redundancy across numerous frames and can comprise thousands of individual frames over a short duration, medical 3D imaging typically consists of fewer slices along the z-axis, usually ranging from 20 to 300. This fundamental difference presents challenges for the application of fixed-size 3D patch embedding methods. Approaches using fixed-size patches, such as those with dimensions of 16 by 16 by 16, often struggle to capture fine anatomical details effectively due to the lower number of slices in the z-axis compared to the high in-plane resolution typical of medical imaging, such as 512 by 512 for CT scans. As a result, while fixed embedding methods have proven effective for processing natural 3D data, they are less suited to the structural and informational characteristics of medical 3D imaging.

Despite these challenges, progress has been made. RadFM~\citep{wu2023towards} supports both 2D and 3D images, focusing on text generation tasks like report generation, rationale generation and VQA. Another significant contribution is CT-CLIP~\citep{hamamci2024foundation}, introducing the open-source dataset CT-RATE, which consists of over 25,000 chest CT volume-report pairs. CT-CLIP demonstrates that contrastive learning can align CT images with reports using a 3D vision encoder and text encoder, similar to CLIP. Both CT2Rep~\citep{hamamci2024ct2rep} and 3D-CT-GPT~\citep{chen20243d} employ the 3D vision encoder architecture from CT-CLIP as part of its chest CT report generation model. Unlike CT2Rep that leverages a simple transformer decoder for report generation only, 3D-CT-GPT leverages LLM and perform both report generation and VQA. The M3D-LaMed model~\citep{bai2024m3d} further advances the field by integrating visual information into LLM through a perceiver module, enabling tasks like captioning, VQA, and segmentation. However, M3D-LaMed still relies on a 3D Vision Transformer (ViT)~\citep{dosovitskiy2020image} with fixed patch embedders, limiting its ability to capture detailed visual information, and retaining challenges in representing complex 3D medical structures effectively.

\begin{figure}[htbp]
    \centering
    \includegraphics[width=\textwidth]{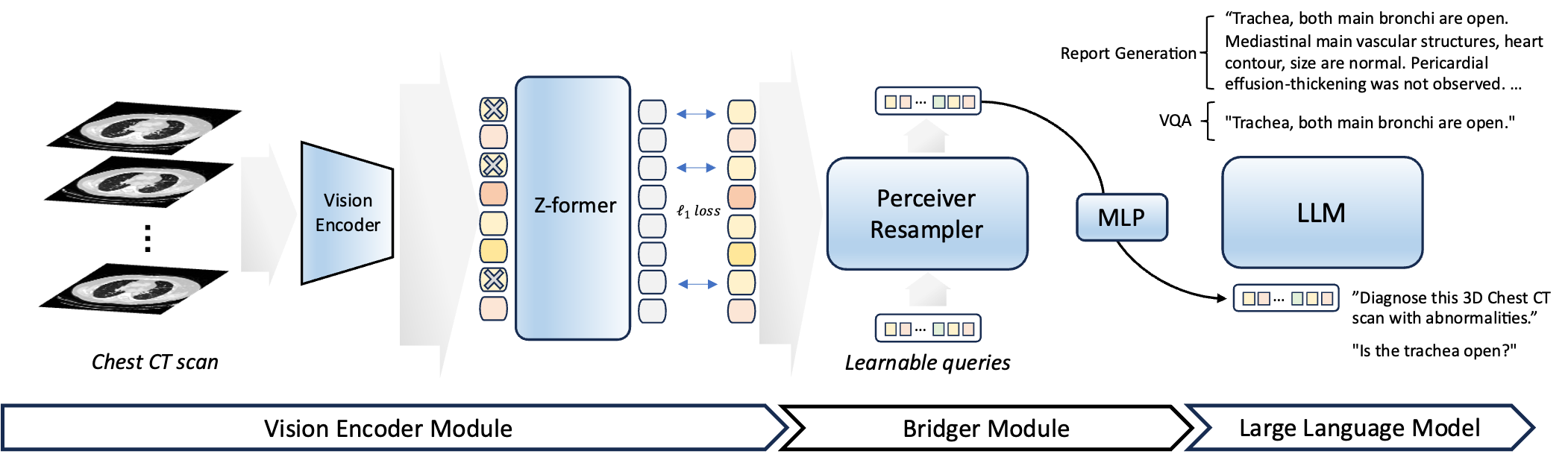} 
    \caption{Overview of MS-VLM architecture.}
    \label{fig:overview}
\end{figure}

\section{Method}

\subsection{Model Architecture}
MS-VLM comprises a vision encoder module, a bridger module, and a LLM as shown in Figure~\ref{fig:overview}. The vision encoder includes a 2D ViT and a novel {\em Z-former} that integrates slice-by-slice 2D ViT embeddings, while the bridger module features a perceiver resampler and an MLP layer to project volumetric representation to a fixed-size learnable query. The learnable query is then used as a visual prompt when instruction fine-tuning the LLM to perform vision-language tasks (report generation or VQA).

Given a 3D medical image, each slice along the z-axis is processed by a pre-trained 2D ViT. The [CLS] token embeddings from the last layer are then concatenated to form a volumetric representation of the given image as a sequence of slice-specific features. This volumetric representation is input to Z-former,  which is constructed using a conventional transformer encoder architecture that employs sparse attention mechanism, Big Bird~\citep{zaheer2020big}. The last hidden layer of Z-former then outputs a volumetric representation as a sequence of sub-volumetric features of which each feature incorporate information from both the original and neighboring slice-specific features. The last hidden layer output of Z-former is then used as keys and values for the perceiver resampler. The learned queries from the perceiver resampler then become the final volumetric representation of the given image and is provided as a visual prompt when instruction fine-tuning the LLM for report generation or VQA. The perceiver resampler allows the model to handle 3D medical images of varying slice lengths by enabling LLM fine-tuning on a fixed-size learnable query.

\subsection{Model Training}

\begin{figure}[htbp]
    \centering
    \includegraphics[width=\textwidth]{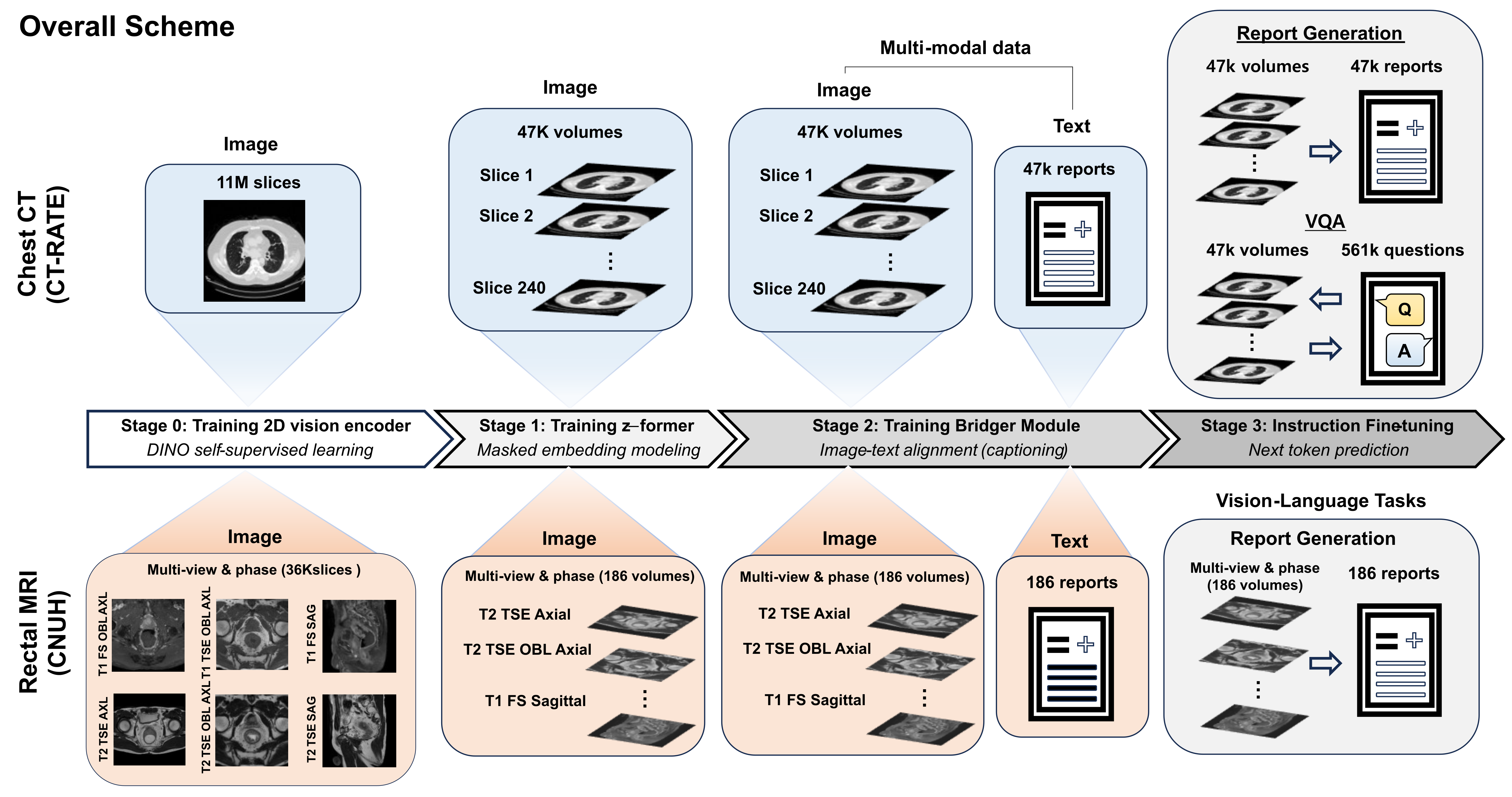} 
    \caption{Overview of MS-VLM training and evaluation. \textbf{a)} MS-VLM is evaluated on public chest CT dataset CT-RATE and in-house Rectal MRI dataset. MS-VLM learns volumetric representation off of a single axial view for chest CT and multi-view for rectal MRI.}
    \label{fig:overall_scheme}
\end{figure}

The training consists of four stages -- Stage 0: Training a domain-specific DINO vision encoder, Stage 1: Training a Z-former, Stage 2: Training a bridger module; Stage 3: Instruction fine-tuning a LLM for report generation and VQA. Figure~\ref{fig:overall_scheme} demonstrates the overall scheme of how the vision and language data are processed for each stage of training. 

\subsubsection{Training a Domain-Specific DINO Vision Encoder}
We utilize a 2D ViT-B/16 model pre-trained on ImageNet-1K using DINO (Self-Distillation with No Labels)~\citep{caron2021emerging} to extract slice-specific spatial features from volumetric images. DINO is a self-supervised learning method that leverages a teacher-student network structure to learn visual representations without labels. By aligning information between views, DINO effectively captures high-quality features suitable for various downstream tasks. During the self-supervised learning process, DINO enables the 2D ViT to learn semantic representations and essential patch-level features, which are crucial for visual understanding. 
For a volumetric image with slice length $L$, each 2D slice $i$ is processed through the pre-trained encoder. We extract the [CLS] token embedding $\mathbf{z}_i^{\text{[CLS]}}$ from the final layer, representing slice-specific spatial features. These embeddings are concatenated along the z-axis to create a comprehensive volumetric representation.
\[
\mathbf{Z}_{\text{vol}} = [\mathbf{z}_1^{\text{[CLS]}}, \mathbf{z}_2^{\text{[CLS]}}, \dots, \mathbf{z}_L^{\text{[CLS]}}]
\]
The concatenated embeddings \( \mathbf{Z}_{\text{vol}} \) are then input to the Z-former for further processing.

\subsubsection{Training the Z-former}

To train the Z-former, we introduce masked embedding modeling (MEM), inspired by masked image modeling~\citep{xie2022simmim}. Instead of predicting pixel values, MEM regresses on [CLS] token embeddings from \( \mathbf{Z}_{\text{vol}} \), where each embedding is masked with a fixed probability. The Z-former reconstructs the masked embeddings, optimizing via $\ell_1$ loss against the original:

\begin{equation}
\label{eq:mem}
\mathcal{L}_{\text{MEM}} = \frac{1}{|M|} \|\mathbf{\hat{Z}}_{\text{masked}} - \mathbf{Z}_{\text{vol}}\|_1
\end{equation}
Here, $M$ refers to the set of masked embeddings and $|M|$ denotes its cardinality. 
The $l_1$ norm is defined as the sum of the absolute differences of each vector embedding.

Using sparse attention, the Z-former captures dependencies along the $z$-axis, effectively learning the relational structure across slices to represent spatial dependencies. The final hidden state embedding encodes sub-volumetric features incorporating information from the current and neighboring slices.

\subsubsection{Training the Bridger Module}

The task-agnostic volumetric representation generated by the Z-former requires further alignment with the language space of the LLM. To address this modality gap, we utilize a perceiver resampler~\citep{alayrac2022flamingo} and an MLP with ReLU activation. The Z-former's final hidden state is projected as keys and values for the perceiver resampler's cross-attention module, where the learnable query attends to specific sub-volumetric representations and aggregates them into refined visual features. These features are subsequently mapped to the LLM's conceptual language space using the MLP, which adjusts the query dimension to match that of the LLM embedding space.

During this process, the DINO vision encoder, Z-former and LLM remain frozen, ensuring the preservation of pre-trained knowledge in the vision encoder module while avoiding catastrophic forgetting in the LLM. The bridger module is optimized to align the visual feature to the LLM embedding space such that aligned feature generate the corresponding reports. The bridger module thus plays a critical role in transforming the visual features into a format that is fully compatible with the LLM’s language space. 

\subsubsection{Instruction Fine-tuning}

\begin{figure}[!t]
    \centering
    \includegraphics[width=1.0\textwidth]{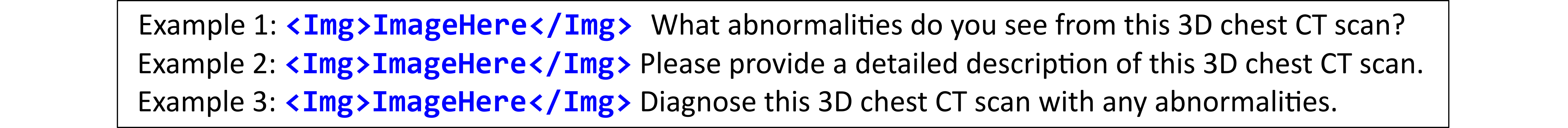}
    \caption{Instruction prompts used for instruction fine-tuning the LLM for report generation.}
    \label{fig:report-prompt}
\end{figure}

We employ Vicuna-7B-v1.5~\citep{chiang2023vicuna} as our LLM, selected for its exemplary instruction-following capabilities among publicly accessible checkpoints~\citep{liu2024visual, chiang2023vicuna, taori2023stanford}. The instructions provided for the chest CT report are illustrated in Figure~\ref{fig:report-prompt}.

The placeholder \texttt{\textless ImageHere\textgreater} represents the learned query generated by the bridger module, which will henceforth be referred to as the visual features. These visual features are encapsulated within \texttt{\textless Img/\textgreater} and \texttt{\textless /Img\textgreater} token embeddings to distinguish them from the report generation instructions. During this stage, the DINO vision encoder and the Z-former remain frozen, while only the bridger module and the LLM are subject to fine-tuning. Specifically, the LLM is fine-tuned using Low-Rank Adaptation (LoRA) by minimizing the negative log-likelihood (NLL) of generating the report $Y =\{ y_1, y_2, \dots, y_T \}$, conditioned on the visual features $Q$ and the instruction $I$.

\begin{equation}
\label{eq:inst}
\mathcal{L}_{\text{inst}} = - \sum_{t=1}^{T} \log P(y_t \mid y_{<t}, Q, I)
\end{equation}

\begin{figure}[!t]
    \centering
    \includegraphics[width=1.0\textwidth]{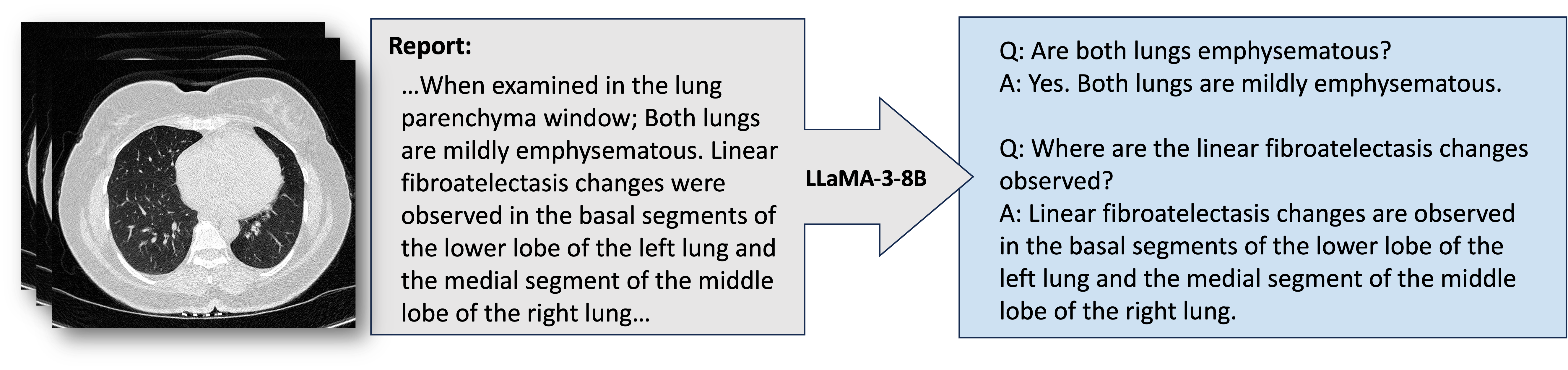}
    \caption{Generating synthetic VQA pairs from CT-RATE reports using LLaMA-3-8B.}
    \label{fig:syn-vqa}
\end{figure}

One key challenge in aligning 3D medical images, such as CT scans, with radiology reports is the presence of multiple findings within a single scan. Radiology reports inherently describe diverse findings, complicating effective alignment between the image and the text, often resulting in inadequate representation of individual findings.

To address this challenge, we decompose radiology reports into individual components and construct a VQA dataset. Each VQA pair specifically queries the presence and attributes (e.g., location or severity) of a single finding, thereby ensuring precise alignment between each distinct finding in the image and its corresponding textual description.
For this purpose, we utilize the RadGenome-ChestCT dataset~\citep{zhang2024radgenome}, which provides VQA pairs related to the presence, location, size, and type of abnormalities in CT-RATE chest CT scans. Additionally, we generate synthetic VQA pairs based on CT-RATE reports using LLaMA-3-8B, as illustrated in Figure~\ref{fig:syn-vqa}, with detailed prompt instructions available in~\ref{app1}. Both the RadGenome-ChestCT dataset and the synthetic VQA pairs are employed in the VQA task.

Finally, MS-VLM is fine-tuned on both the report generation task and the VQA task. The report generation task aims to generate comprehensive radiology reports, while the VQA task focuses on answering specific questions about individual findings. These tasks are jointly fine-tuned with a 1:1 sample ratio per training iteration.

\subsection{Implementation Details}

\subsubsection{Datasets}

For the chest CT report generation task, we evaluate our model using the CT-RATE dataset, introduced in CT-CLIP~\citep{hamamci2024foundation}. The CT-RATE dataset comprises 25,692 non-contrast chest CT volumes from 21,304 unique patients. With multiple reconstructions, the dataset expands to 50,188 volumes, each paired with radiology reports detailing findings and impressions. 

We train our model on 47,149 volumes from 20,000 unique patients, excluding cases with insufficiently short reports. The test set consists of 3,039 volumes from 1,304 unique patients. To ensure a fair comparison with existing methods, such as CT2Rep, we focus exclusively on the findings sections of the reports.

CT volumes are preprocessed following the CT-CLIP protocol~\citep{hamamci2024foundation}, which includes resizing to achieve uniform spacing of 1.5 mm along the z-axis and 0.75 mm along the x- and y-axes. Each volume is then padded or center-cropped to maintain a consistent resolution of \( 480 \times 480 \times 240 \) across the dataset.

For the chest CT VQA task, we utilize both synthetic VQA pairs and pairs from the RadGenome-ChestCT dataset~\citep{zhang2024radgenome}. The synthetic dataset, comprising approximately 180,000 QA pairs, excludes instances with overly brief answers, such as ``Yes" or ``No." RadGenome-ChestCT, a large-scale dataset with detailed 3D chest CT annotations, provides 1.3 million QA pairs based on CT-RATE. For training, we filter QA pairs to retain only those mentioning the most specific anatomical regions associated with common abnormalities, ensuring precise alignment. An example QA pair is provided in~\ref{app1}. For training, we randomly select 121,000 QA pairs related to the types of abnormalities, 136,000 presence-related QA pairs, 110,000 location-related QA pairs, and 14,000 size-related QA pairs, creating a comprehensive training dataset that ensures robust coverage across different aspects of chest CT interpretation.

\begin{figure}[!t]
    \centering
    \includegraphics[width=1.0\textwidth]{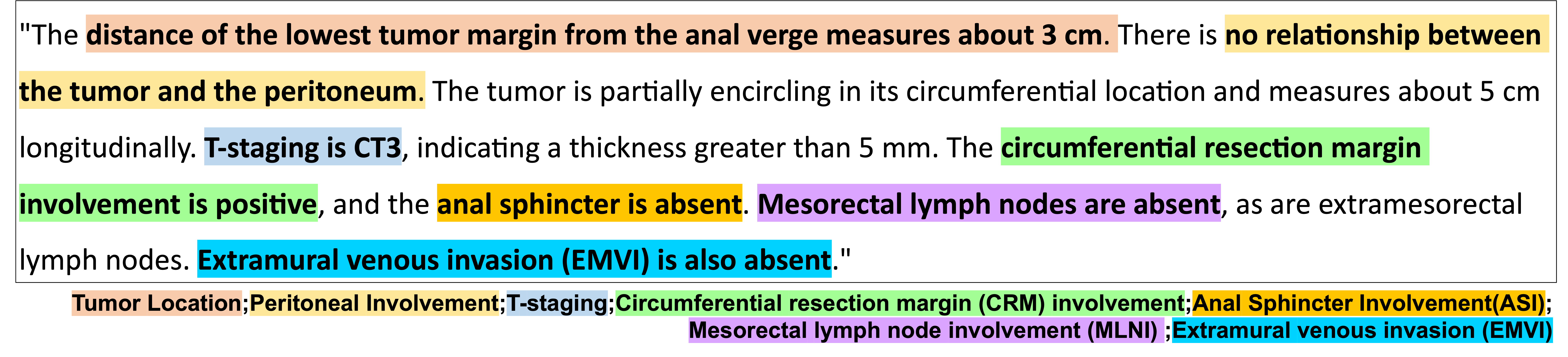}
    \caption{An example of a rectal MRI report that describes the seven most common rectal tumor-related abnormalities. Each sentence is color-coded based on the abnormality it describes.}
    \label{fig:rectal-report}
\end{figure}

The rectal MRI report generation was conducted using in-house data obtained from Chungnam University Hospital. The dataset comprised MRI volumes from 311 patients and their corresponding radiology reports. Most of the radiolgy reports, though unstructured and presented in free-form text, consistently contained information on seven key findings: tumor location, peritoneal involvement, T staging, circumferential resection margin (CRM) involvement, anal sphincter involvement (ASI), mesorectal lymph node involvement (MLNI), and extramural venous invasion (EMVI). Tumor location was measured in centimeters from the anal verge to guide treatment decisions, while peritoneal involvement was categorized as none, partial, or full to assess the extent of invasion. T staging classified the tumor's depth (T1-T4), indicating its local invasiveness. The CRM was evaluated to determine the proximity of the tumor to the mesorectal fascia, which serves as an indicator of local recurrence risk. ASI assessed the potential tumor extension into the anal sphincter, influencing surgical options regarding sphincter preservation. MLNI involved checking for signs of lymph node enlargement or metastasis, which are crucial prognostic factors. EMVI assessed the tumor’s invasion into extramural veins, which is associated with an increased risk of metastasis (refer to Figure~\ref{fig:rectal-report} for details). 

The MRI volumes for each patient were acquired using three different contrasts: T1-weighted (T1), T2-weighted (T2), and Diffusion-Weighted Imaging (DWI). The T1 and T2 volumes were obtained across five imaging planes: axial, oblique axial, sagittal, coronal, and oblique coronal, while the DWI volumes were captured exclusively in the axial plane. The T1 volumes were acquired with or without modifiers such as Fat Suppression (FS), Spectral Presaturation with Inversion Recovery (SPIR), and Contrast Enhancement (CE) to improve image clarity. For some patients, both modified and unmodified versions of T1 volumes were available. Overall, the MRI volumes per patient included at least four axial volumes, two sagittal volumes, and one coronal volume.

These multi-view and multi-phase MRI volumes were utilized to compare a baseline model that exclusively uses T2 axial volumes, which are considered most critical for rectal tumor MRI interpretation, with the proposed MS-VLM, capable of robustly processing all imaging phases regardless of sequence length. Further details regarding the availability of volumes for each imaging plane and contrast phase can be found in~\ref{app2}.

The study received approval from the Institutional Review Board (IRB) of Chungnam University Hospital (IRB number 2022-05-113), and the requirement for informed consent was waived due to the retrospective nature of the study.

\subsubsection{Training Details}
To train on CT-RATE during Stage 0, we employed a ViT-B/16 model as the DINO vision encoder, initialized with ImageNet-1K DINO pretrained weights. We fine-tuned the model for 50 epochs on volumetric image slices using DINO. Global and local crops were applied with area ranges of 0.8–1.0 and 0.1–0.3, respectively, relative to the input image area. Data augmentations included random rotations, auto-contrast, and equalization.

For subsequent training stages, input image dimensions were set to \(240\times 3\times 480\times 480\), representing full volumes. A Z-former, with 12 hidden layers and a model dimension of 768, was employed. The Z-former utilized a sparse attention mechanism with a sliding window size of 16 and three random blocks, without global blocks. It was pre-trained for 20 epochs using masked embedding modeling (MEM) to capture dependencies along the z-axis. The masking probability was set to 0.3. Both the 2D ViT and Z-former were trained with the Adam optimizer with learning rate of 1e-4 and an initial warmup phase of 50 linear iterations starting from 1e-5.

The perceiver resampler was initialized with 32 learnable queries, each with a dimension of 4096. Both the perceiver resampler and an associated MLP layer were randomly initialized before training. For the LLM, Vicuna-7B-1.5v was used. The bridger module was trained for 1 epoch, followed by 5 epochs of joint training with the LLM. An AdamW optimizer was used with a learning rate of 1e-4 and a weight decay of 0.05. 

The rectal MRI dataset consisted of 311 patients, of which 82 were excluded from training due to radiology reports lacking sufficient detail on the seven key findings. The remaining dataset was divided into 186 patients for training and 43 patients for testing.
For all training stages, we employed a configuration similar to the CT-RATE training pipeline. Specifically, a domain-specific DINO vision encoder was pre-trained in a self-supervised manner using only the T1/T2 axial and sagittal views, as these are the most critical planes for analyzing rectal tumors.
In cases where patients had more volumes than the required number, a prioritization strategy was applied to select volumes in the following order: T2 volumes, T1 volumes with modifiers (e.g., FS, CE), and T1 volumes without modifiers. This order reflects the relative significance of these imaging modalities in diagnosing rectal tumors.
As a result, 20 slices were sampled from each of the six selected phases, leading to a stacked input configuration of 
\(120\times 3\times 480\times 480\).

All experiments on CT-RATE were conducted on eight NVIDIA A100 (40GB) GPUs, while those on rectal MRI were conducted on a single NVIDIA A100 (40GB) GPU.

\subsection{Evaluation Metrics for Chest CT}

To assess the clinical efficacy of the generated radiology reports, we employ two evaluation approaches: natural language generation (NLG) metrics, including BLEU~\citep{papineni2002bleu}, ROUGE~\citep{rouge2004package}, and METEOR~\citep{lavie2009meteor}, and clinical accuracy (CA) metrics, including precision, recall and F1, as utilized in CT2Rep. NLG metrics quantify the linguistic similarity between the generated reports and the ground truth by evaluating n-gram overlaps, sequence alignment, and synonym usage. These metrics provide an objective measure of the overall linguistic quality and coherence of the generated reports with respect to the ground truth.

For CA metrics, we use a RadBERT-based text classifier introduced in CT-CLIP~\citep{yan2022radbert, hamamci2024foundation} to determine whether the generated reports include the abnormalities present in the ground truth. We calculate precision, recall, and F1 scores for each abnormality to evaluate this alignment. CA metrics thus measure how well the generated reports capture critical findings from the ground truth while minimizing hallucinated content.

Radiology reports may differ in specific phrasing or structure while conveying fundamentally the same information. To accommodate such variability, we employed an additional evaluation framework utilizing LLMs. Conventional NLP metrics, which primarily capture structural similarity, often fail to account for semantic equivalence. Similarly, CA metrics provide quantifiable measures but struggle to reflect nuanced or alternative expressions of equivalent clinical content. LLM-based evaluation offers a potential solution by addressing these limitations and enabling more flexible and context-aware assessments. 

In the evaluation of chest CT reports, four key categories were defined: Presence, Severity, Location, and Hallucination. Each category is assessed binarily, with a score of 1 indicating complete accuracy across all abnormal findings in the reference report. A score of 0 signifies the presence of factual errors in the generated report concerning the relevant category. GPT-4o-mini was used for evaluation. Detailed evaluation prompts for each category are provided in~\ref{app3}.

\subsection{Evaluation Metrics for Rectal MRI}
The evaluation of rectal MRI involved the seven key findings previously mentioned to assess tumor staging and invasion: tumor location, peritoneal involvement, T staging, CRM involvement, ASI, MLNI, and EMVI. Given the heterogeneous nature of these findings, distinct evaluation approaches were employed based on their characteristics.

Metrics such as tumor location and peritoneal involvement are inherently complex and difficult to evaluate using straightforward binary classification. These findings often involve nuanced descriptions and categorizations, such as precise tumor distance from the anal verge or the extent of peritoneal invasion. To address this complexity, a LLM was utilized to indirectly assess the validity of these metrics. By leveraging advanced contextual understanding, the LLM provided an approximate measure of accuracy. Specifically, o1-mini was employed for this evaluation, enabling a flexible and context-aware assessment of these findings. The prompts used for o1-mini evaluation is detailed in~\ref{app3}.

In contrast, findings such as CRM involvement, ASI, MLNI, and EMVI, which can be distinctly classified as either positive or negative, were evaluated using CA metrics. These metrics quantified the classification accuracy through measures such as precision, recall, and F1 scores, providing a robust evaluation of the generated reports' ability to accurately reflect the presence or absence of these clinically significant features.

\section{Experimental Results}

\subsection{Experimental Results on CT-RATE}
Among the existing methods for chest CT report generation, we selected CT2Rep and 3D-CT-GPT as primary baselines due to their demonstrated effectiveness in processing chest CT scans. Other approaches, such as RadFM and M3D-LaMed, rely on pre-training datasets sourced from Radiopedia, and their performance on report generation has been shown to lag behind that of 3D-CT-GPT. Accordingly, our evaluation emphasizes models that are specialized for chest CT report generation. 

In addition, to demonstrate the efficiency of MS-VLM's volumetric representation processing, we conducted an additional experiment where the vision encoder in MS-VLM was replaced with a 3D ViT, closely mirroring the architecture used in existing models. We refer to this MS-VLM variant as MS-VLM (3D encoder). 

Only stages 2 and 3 of the MS-VLM training procedure are applied to training MS-VLM (3D encoder), with the 3D ViT also being trained for both stages. A direct comparison with MS-VLM (3D encoder) under identical experimental conditions enables a clear evaluation of the improvements introduced by our approach.

Using the official implementation and training settings provided for CT2Rep, we successfully reproduced its results and evaluated its performance on the CT-RATE test set. Since the code for 3D-CT-GPT is not publicly available, we relied on the performance metrics reported in the original paper~\citep{chen20243d}. Specifically, we compared against the 3D-CT-GPT (T2) model, which was pre-trained on a private chest CT dataset and validated on CT-RATE. The results on 3D-CT-GPT that is trained and validated solely on CT-RATE were unavailable.

\subsubsection{Report Generation Results} 

\begin{table}
    \centering
    \resizebox{1.0\textwidth}{!}{%
    \begin{threeparttable}
    \caption{\\
    Comparison of NLG metrics and CA metrics on the generated reports between MS-VLM and baseline methods.}
    \label{tab:nlg_ce_metrics}
    \begin{tabular}{lcccccccc}
    \toprule
    \textbf{Method} & \multicolumn{3}{c}{\textbf{NLG Metrics}} & \multicolumn{3}{c}{\textbf{CA Metrics}} \\
    \cmidrule(lr){2-4} \cmidrule(lr){5-7}
     & \textbf{BLEU-4} & \textbf{ROUGE-L} & \textbf{METEOR} & \textbf{Precision} & \textbf{Recall} & \textbf{F1 Score} \\
    \midrule
    3D-CT-GPT & 0.133 & 0.145 & 0.140 & - & - & - \\
    CT2Rep (3D encoder) & \textbf{0.252} & \underline{0.432} & 0.328 & \textbf{0.355} & 0.132 & 0.175 \\
    MS-VLM (3D encoder) & 0.184  & 0.364 & \underline{0.331} & 0.193 & 0.231 & 0.207 \\
    MS-VLM (w/o VQA) & 0.189 & 0.374 & 0.313 & 0.192 & \underline{0.298} & \underline{0.232} \\ 
    MS-VLM (proposed) & \underline{0.232} & \textbf{0.438} & \textbf{0.396} & \underline{0.222} & \textbf{0.329} & \textbf{0.261} \\ 
    \bottomrule
    \end{tabular}
    \begin{tablenotes}\footnotesize
    \item[*] The best performance is highlighted in bold, and the second-best is underlined. CA metrics are averaged across 18 chest-related abnormalities.
    \end{tablenotes}
    \end{threeparttable}
    }
\end{table}

\begin{figure}
    \includegraphics[width=1.0\textwidth, height=0.9\textheight, keepaspectratio]{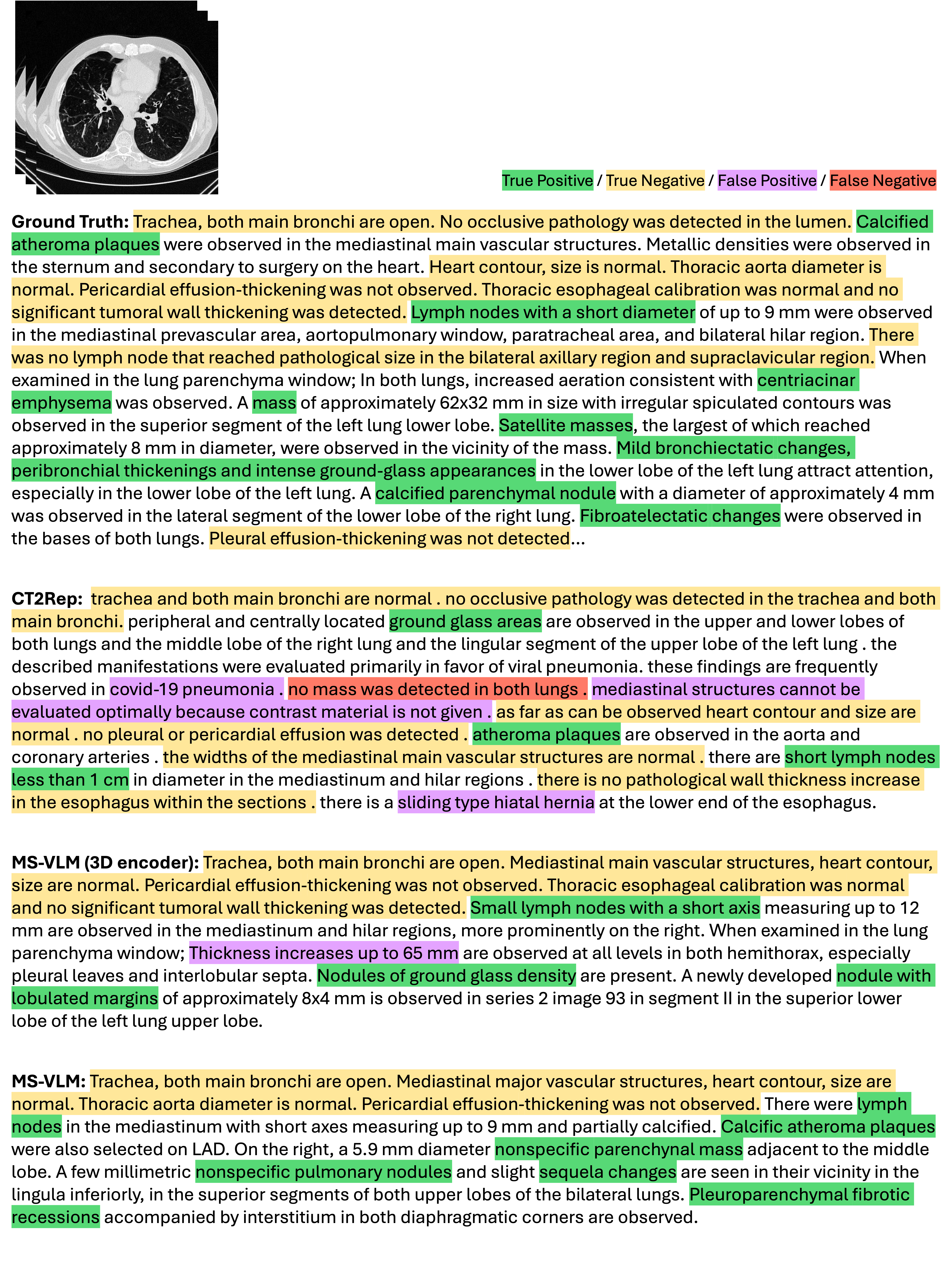}
    \centering
    \caption{Qualitative evaluation on CT-RATE. Each sentence from the report is color coded to identify True Positive, True Negative, False Positive and False Negatives on the findings mentioned in the ground truth report. MS-VLM achieves greater recall on the true positives and is less prone to generating false negatives. MS-VLM (3D encoder): A variant of MS-VLM using 3D ViT encoder without Z-former.}
    \label{fig:qualitative-example1}
\end{figure}

As presented in Table~\ref{tab:nlg_ce_metrics}, MS-VLM demonstrates superior performance in radiology report generation compared to baseline methods. It consistently surpasses 3D-CT-GPT (T2) and CT2Rep in ROUGE and METEOR metrics, highlighting its ability to generate reports that closely align with the original radiology report structure annotated by experts.

In terms of CA metrics, MS-VLM achieves the highest overall performance, with an average F1 score of 0.261. While CT2Rep demonstrates higher precision, its lower recall reduces its overall F1 score. The qualitative comparison in Figure~\ref{fig:qualitative-example1} further highlights the superiority of MS-VLM. It effectively recalls a greater number of abnormal findings, generating reports that closely align with the ground truth. Notably, MS-VLM demonstrates the ability to identify subtle positive findings, such as satellite masses and fibrotic changes, which are often overlooked by other baseline models. 

MS-VLM also outperforms its variant, MS-VLM (3D encoder), which is a variant of MS-VLM using 3D ViT encoder without Z-former, in both NLG and CA metrics. Higher average ROUGE, METEOR, and F1 scores confirm that MS-VLM generates reports that better match radiologists’ annotations while addressing critical findings more effectively, demonstrating the utility of the proposed combination of a 2D vision encoder and the Z-former compared to conventional 3D vision encoder approaches. As shown in Figure~\ref{fig:qualitative-example1}, MS-VLM identifies more true positives compared to MS-VLM (3D encoder). For additional qualitative analyses, refer to~\ref{app4}. Similar trends are observed in comparisons with other MS-VLM variants. When VQA task is not performed jointly with report generation task during training, MS-VLM (w/o VQA) performance becomes suboptimal, suggesting the importance of integrating both tasks for optimal performance.

\begin{table}
    \centering
    \resizebox{0.85\textwidth}{!}{%
    \begin{threeparttable}
    \caption{\\ 
    Comparison of GPT-4o-mini evaluation performance between MS-VLM and baseline methods.}
    \label{tab:gpt4omini_metrics}
    \begin{tabular}{lcccc}
    \toprule
    \textbf{Method} & \textbf{Presence} & \textbf{Location} & \textbf{Severity} & \textbf{Hallucination} \\
    \midrule
    CT2Rep (3D encoder) & 0.186 & 0.197 & 0.216 & \textbf{0.186} \\
    MS-VLM (3D encoder) & 0.203 & \underline{0.216} & \underline{0.223} & 0.122 \\
    MS-VLM (w/o VQA) & \underline{0.212} & 0.207 & 0.209 & 0.124 \\ 
    MS-VLM (proposed) & \textbf{0.217} & \textbf{0.225} & \textbf{0.234} & \underline{0.127} \\ 
    \bottomrule
    \end{tabular}
    \begin{tablenotes}\footnotesize
    \item[*] The best performance is highlighted in bold, and the second-best is underlined.
    \end{tablenotes}
    \end{threeparttable}
    }
\end{table}

For quantitative evaluation, detailed CA metric results for MS-VLM across 18 abnormality categories are provided in~\ref{app5}.
Table~\ref{tab:gpt4omini_metrics} presents the evaluation results using GPT-4o-mini. Unlike abnormality-specific evaluations, GPT-4o-mini assesses the overall report’s accuracy for each category on a binary scale. Due to this strict evaluation criterion, even minor inaccuracies result in a score of zero, leading to relatively low average scores across all categories for all models.

MS-VLM demonstrates superior performance in the presence, location, and severity categories, with the exception of the hallucination category. These results indicate that MS-VLM not only accurately predicts abnormalities but also exceeds baseline models in identifying their precise location and severity. However, in the hallucination category, MS-VLM achieves slightly lower scores compared to CT2Rep. 
The relatively better performance of CT2Rep in the hallucination category is likely attributable to its tokenizer, which has been specifically trained on CT-RATE radiology reports.

MS-VLM outperforms both its variants, MS-VLM (3D encoder) and MS-VLM (w/o VQA), across all evaluated categories. These results suggest the superiority of the proposed Z-former-based approach over traditional 3D vision encoder methods. Furthermore, they highlight the critical role of VQA training focused on single abnormalities in improving the model's understanding and alignments for the fine-grained findings, thereby enabling more precise attention to detailed abnormalities and enhancing overall model performance.

\subsubsection{Robustness of Z-former}

\begin{table}
    \centering
    \resizebox{0.66\textwidth}{!}{%
    \begin{threeparttable}
    \caption{\\
    Comparison of CA metrics of the generated reports when MS-VLM is given volumes with fixed z-length versus volumes with original slice length.} 
    \label{tab:method_comparison}
    \begin{tabular}{lccc}
    \toprule 
    \textbf{Method}        & \textbf{Precision}    & \textbf{Recall}    & \textbf{F1}   \\ 
    \midrule
    \textbf{all volumes} \\
    CT2Rep ($z=240$)                & \textbf{0.355}         & 0.132         & 0.175         \\
    MS-VLM ($z=240$)         & 0.222         & \textbf{0.329}         & \textbf{0.261}         \\
    MS-VLM ($z=L$)    & \underline{0.234}  & \underline{0.286} & \underline{0.245} \\ 
    \midrule
    \textbf{volumes with $L>240$} \\
    CT2Rep ($z=240$) & \textbf{0.330} & 0.146 & 0.190 \\ 
    MS-VLM ($z=240$)       & \underline{0.229}         & \underline{0.258}         & \underline{0.237}         \\
    MS-VLM ($z=L$)       & 0.226         & \textbf{0.284}         & \textbf{0.246}         \\
    \bottomrule
    \end{tabular}
        \begin{tablenotes}\footnotesize
    \item[*] $z$ denotes the number of slices in z-axis, and $L$ denotes original slice length of the volume.
    \end{tablenotes}
    \end{threeparttable}
    }
\end{table}

Our model inherently integrates visual information into a single [CLS] token embedding $\mathbf{z}_i^{\text{[CLS]}}$ irrespective of the z-length, enabling it to process visual information seamlessly across varying z-lengths. This capability offers a significant advantage in clinical settings where slice lengths are not fixed, as scan ranges are determined post-scout view (tomogram) acquisition, resulting in variable slice lengths for each patient. This flexibility contrasts with traditional models that rely on 3D vision encoder with a fixed z-length, which may lead to information loss. 

To validate this hypothesis, we conducted experiments analyzing the impact of slice count on CT volume performance. For a fair comparison with previous studies, the baseline setting involves fixing the z-length to a constant value of 240. However, when the z-length $L$ exceeds 240, fixed z-length approaches may suffer from significant information loss. We compared our method, MS-VLM which uses original slice length without fixing the z-length, against both traditional fixed z-length methods and MS-VLM with fixed z-length.

Table~\ref{tab:method_comparison} summarizes the experimental results. The MS-VLM approach, which retains the original slice length without imposing a fixed z-length, exhibits no significant performance degradation, demonstrating its adaptability. 
This advantage is particularly evident in subgroup analyses of data with longer slice length ($L > 240$), where the proposed method outperforms not only fixed z-length approaches such as CT2Rep but also the fixed z-length variant of MS-VLM. These findings effectively validate our hypothesis and suggest the utility of MS-VLM in clinical scenarios involving variable z-lengths.

\subsection{Experimental Results on Rectal MRI}

\begin{table}
    \centering
    \resizebox{1.0\textwidth}{!}{%
    \begin{threeparttable}
    \caption{\\
    Comparison of o1-mini evaluation performance and F1 scores of abnormality prediction between MS-VLM and baseline methods.}
    \label{tab:rectal-combined}
    \begin{tabular}{lcccccccccc}
    \toprule
    \textbf{Model} & \textbf{Location} & \textbf{Peritoneum} & \textbf{T Stage} & \textbf{CRM} & \textbf{ASI} & \textbf{MLNI} & \textbf{EMVI} \\ \hline
                   & \multicolumn{3}{c}{\textbf{Accuracy}} & \multicolumn{4}{c}{\textbf{F1}} \\
    \cmidrule(lr){2-4} \cmidrule(lr){5-8}
    CT2Rep (T2 axial) & \underline{0.209} & \underline{0.488} & 0.163 & 0.143 & \underline{0.133} & 0.108 & 0.000 \\
    MS-VLM (3D encoder, T2 axial) & 0.116 & 0.372 & 0.116 & 0.171 & 0.121 & \underline{0.426} & \underline{0.111} \\
    MS-VLM (T2 axial) & \underline{0.209} & 0.395 & \underline{0.279} & \underline{0.308} & 0.114 & 0.298 & \underline{0.111} \\
    MS-VLM (all views \& phases) & \textbf{0.233} & \textbf{0.512} & \textbf{0.581} & \textbf{0.421} & \textbf{0.222} & \textbf{0.667} & \textbf{0.143} \\
    \bottomrule
    \end{tabular}
    \begin{tablenotes}\footnotesize
    \item[*] {Accuracy is evaluated using o1-mini. The best values are highlighted in bold, and the second-best are underlined. The name of the abnormalities are shortened for readability.}
    \end{tablenotes}
    \end{threeparttable}}
\end{table}

The rectal MRI experiments were conducted to evaluate the proposed MS-VLM architecture under key conditions characterized by fewer slices per volume compared to CT, the necessity of integrating multiple views (e.g., axial, sagittal) and phases (e.g., T1, T2), and limited data availability. These conditions were designed to assess the advantages of MS-VLM’s pre-trained 2D vision encoder using DINO and its subsequent Z-former self-supervised learning approach, along with the overall training framework, in efficiently handling multi-phase and multi-view data and adapting to data-scarce clinical settings.

Table~\ref{tab:rectal-combined} summarizes the experimental results, demonstrating that MS-VLM outperformed baseline methods, including CT2Rep and MS-VLM (3D encoder), across all critical findings. Notably, the proposed MS-VLM achieved significant improvements in key metrics such as T stage classification, CRM threatening, and MLNI detection. In these metrics, MS-VLM not only outperformed CT2Rep and MS-VLM (3D encoder) but also exceeded the performance of a variant using only the primary view, T2 axial, within the identical MS-VLM framework.
These results validate MS-VLM’s ability to effectively integrate information from multi-view and multi-phase data, underscoring the importance of such integration for optimal performance. For qualitative examples of the generated reports, refer to~\ref{app4}.

\subsection{Results on Computational Efficiency}
\begin{figure}
    \includegraphics[width=0.75\textwidth]{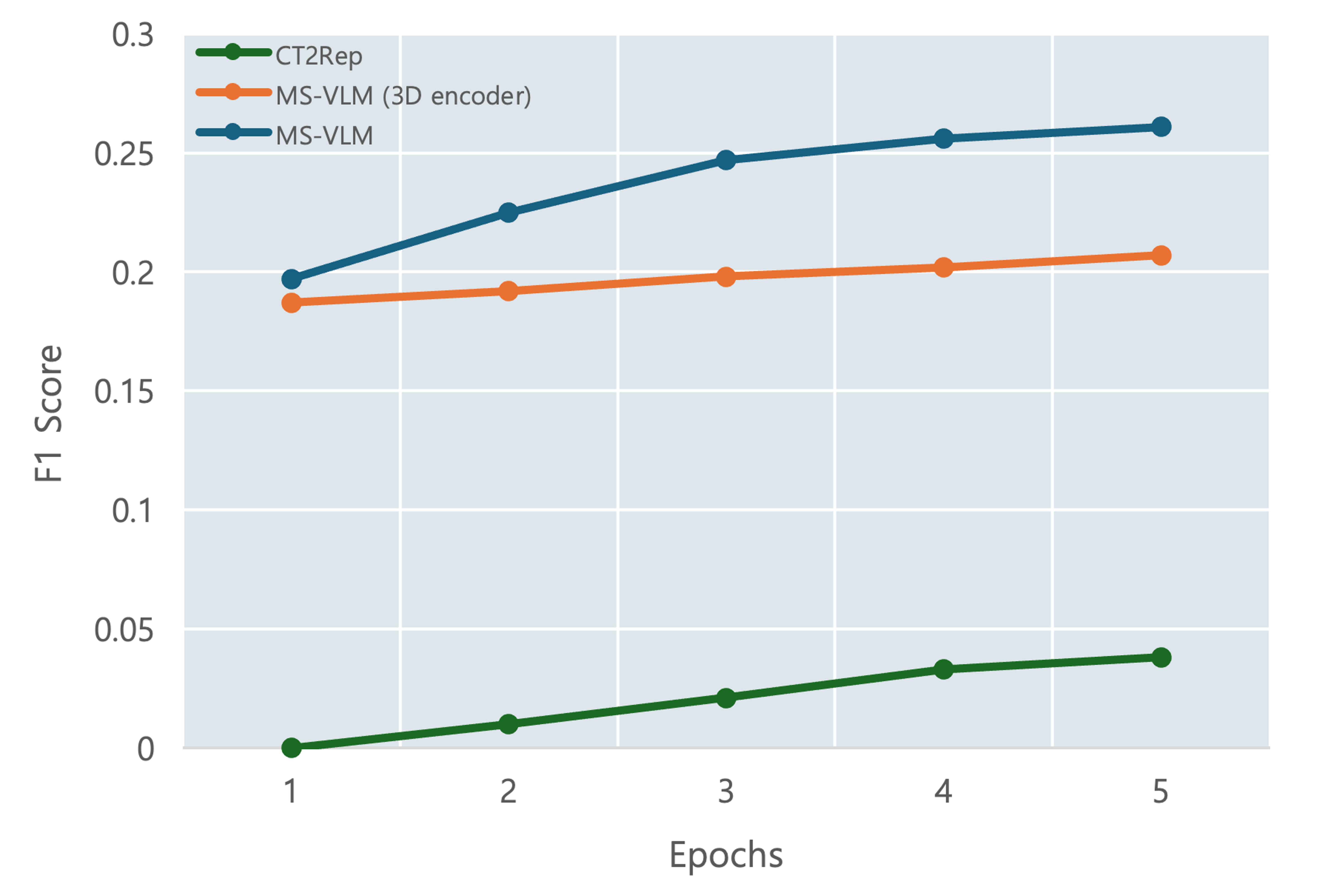}
    \centering
    \caption{Average F1 score across 18 chest abnormalities over training epochs. MS-VLM achieves superior average F1 score compared to baseline methods.}
    \label{fig:eff-graph}
\end{figure}

Our approach leverages DINO to effectively learn critical visual representations from individual slices in the initial stage, followed by the Z-former to capture inter-slice dependencies. This two-step strategy facilitates rapid convergence during subsequent training with 3D volumetric data, offering computational efficiency in terms of data utilization. Furthermore, our method maximizes the benefits of pre-trained LLMs by separating the tuning process into distinct stages, allowing us to exploit the advantages of pre-training to further improve computational efficiency. To validate this hypothesis, we compared the convergence speed of models during the training process.
Figure~\ref{fig:eff-graph} demonstrates that the MS-VLM architecture achieves the fastest convergence in CA metric performance compared to other methods. For instance, CT2Rep attains an average F1 score of only 0.038 after the same five epochs, indicating that it requires significantly more training to achieve comparable performance. Specifically, CT2Rep needed 40 epochs of training to reach an average F1 score of 0.175.
Notably, the proposed MS-VLM approach, which utilizes a 2D vision encoder and the Z-former, achieves faster convergence and superior performance compared to methods employing a 3D vision encoder. This demonstrates the efficiency of the sequential training strategy, wherein DINO is used to pre-train the 2D vision encoder, followed by training the Z-former to capture inter-slice dependencies.
These results suggest the computational advantages of our staged approach, effectively leveraging the strengths of pre-trained LLMs to achieve efficient convergence while maintaining high performance.

\section{Discussion}
Recent advancements in multi-modal AI models and improvements in model scalability have underscored the potential of true multi-modal medical foundation models~\citep{tu2024towards, zhang2024generalist, lu2024multimodal}. These models aim to integrate diverse data modalities such as images and text, with a growing emphasis on extending capabilities beyond 2D images to volumetric 3D imaging modalities like CT and MRI~\citep{wu2023towards, hamamci2024foundation, chen20243d, bai2024m3d}. Despite these promising developments, the performance of VLMs for 3D medical imaging remains suboptimal, particularly in tasks requiring precise alignment of image features with textual descriptions.

A key limitation lies in the relative under-performance of vision encoders compared to LLMs. While LLMs excel in understanding detailed clinical context and generating coherent reports, vision encoders often struggle to embed fine-grained anatomical details from 3D medical images effectively~\citep{hu2024advancing}. This discrepancy arises from the inherent differences in how current models process 3D volumetric data.

Most existing approaches treat 3D medical images in a manner analogous to videos, employing 3D patch embeddings within 3D vision encoders. However, this methodology overlooks critical distinctions between 3D medical imaging and video data. Videos typically consist of a large number of frames (e.g., a 5-minute video at 30 frames per second results in $224 \times 224 \times 9000$), characterized by substantial redundancy across frames, which makes 3D patch embeddings effective for capturing spatial and temporal correlations~\citep{madan2024foundation}. In contrast, 3D medical images generally contain significantly fewer slices (e.g., $512 \times 512 \times 24$) with lower redundancy between adjacent slices. Under such conditions, the use of 3D patch embeddings can lead to over-correlated representations along the z-axis, potentially impairing model performance. This limitation is particularly pronounced when subtle pathologies appear or disappear between slices, making it crucial to avoid excessive correlation that could obscure meaningful slice-specific information.

To address these challenges, we developed MS-VLM, a VLM architecture inspired by radiologists' workflows. Radiologists typically analyze 3D medical images by examining individual slices sequentially and synthesizing information across slices and views. They also focus on inter-slice relationships; for instance, a structure suspected to be a tumor might be identified as a blood vessel if it continues across slices, whereas a nodular shape that disappears in subsequent slices may be classified as a tumor. Radiologists integrate these observations with multi-view (e.g., axial, sagittal) and multi-phase (e.g., T1, T2) imaging to arrive at a comprehensive interpretation.

MS-VLM mirrors this process by independently embedding slice-specific features using a 2D vision encoder and aggregating them into volumetric representations through the Z-former. This approach significantly reduces redundant correlations along the z-axis while preserving slice-specific granularity and inter-slice relationships, thereby enhancing diagnostic accuracy.

The flexibility of MS-VLM was demonstrated in experiments involving varying CT scan ranges. In scenarios with a larger number of slices, MS-VLM outperformed fixed-length models, which often experience information loss when exceeding their predefined length, by effectively leveraging its ability to aggregate meaningful information across the entire volume without such losses. This adaptability is particularly valuable in clinical settings, where scan ranges can vary depending on the target anatomy or the imaging protocol employed. 

Moreover, MS-VLM exhibited robust performance in multi-view and multi-phase integration, both of which are critical for comprehensive diagnostic workflows. For instance, in rectal MRI experiments, MS-VLM achieved superior accuracy by simultaneously processing axial and sagittal views alongside multi-phase data, such as T1, T2 images. This capability underscores its clinical relevance, as the integration of multi-view and multi-phase data is essential for identifying critical abnormalities and ensuring precise diagnoses.

Another significant advantage of MS-VLM is its reliance on 2D vision encoders, which benefit from advanced pre-training techniques like DINO. This design enables the model to converge rapidly and efficiently learn visual representations, demonstrating improved computational efficiency in terms of fast convergence, as shown in our experiments. Furthermore, MS-VLM’s architecture is highly generalizable, capable of adapting to various imaging modalities, including CT, MRI, PET-CT, and 3D ultrasound. This adaptability positions MS-VLM as a versatile tool for diverse clinical applications, ranging from diagnostic imaging to complex multi-modal analyses.

Despite its promising results, MS-VLM has several limitations. First, the model is not entirely free from hallucination errors and may occasionally miss critical findings. While this study introduces a novel proof-of-concept architecture, it is important to note that larger-scale studies utilizing more diverse and higher-quality datasets could potentially achieve better absolute performance. Second, the validation was limited to chest CT and rectal MRI, leaving the model's generalizability to other imaging modalities, such as abdominal CT, unexplored. Further research is necessary to confirm its applicability across a broader range of clinical scenarios. Third, this study did not include direct expert evaluation, which would provide critical insights into the model’s clinical utility and guide its refinement for practical deployment.

\section{Conclusion}
In this paper, we present MS-VLM, which introduces a novel Z-former architecture that integrates slice-by-slice embeddings into a volumetric representation for 3D medical image interpretation. MS-VLM effectively addresses the challenges of processing 3D medical imaging by adopting a radiologist-inspired approach that preserves slice-specific granularity and captures inter-slice relationships. Its ability to integrate multi-view and multi-phase information, combined with flexibility across varying scan ranges, demonstrates its adaptability to diverse clinical scenarios. These advancements position MS-VLM as a significant step forward in advancing medical VLMs, particularly by enhancing vision encoders to process the complexity of 3D medical imaging with broad generalizability across various modalities.

\appendix

\section{Details on Synthetic VQA Generation}
\label{app1}
\setcounter{figure}{0}
\setcounter{table}{0}

Inspired by LLM-CXR~\citep{lee2024llmcxr}, we generate an instruction prompt that guides LLaMA-3-8B to generate synthetic VQA pairs from a radiology report. 

\begin{figure}[H]
    \centering
    \includegraphics[width=0.8\textwidth, height=0.8\textheight, keepaspectratio]{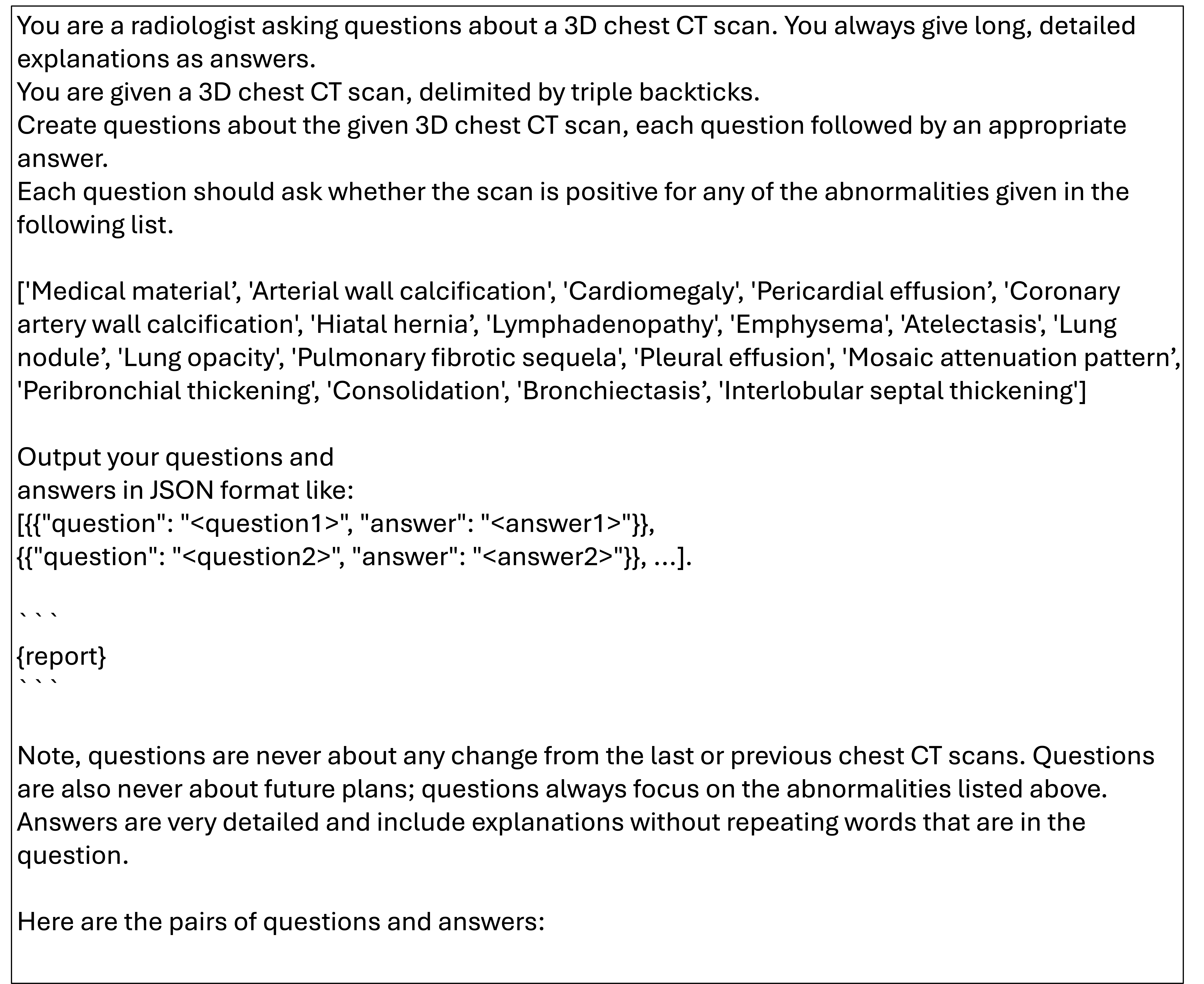}
    \caption{Instruction prompt given to LLaMA-3-8B for synthetic VQA generation from CT-RATE reports.}
    \label{fig:synthetic-vqa-template}
\end{figure}

\begin{figure}[H]
    \centering
    \includegraphics[width=1.0\textwidth]{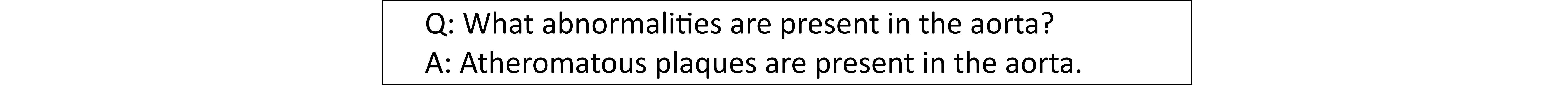}
    \caption{An example of a QA pair from RadGenome-ChestCT dataset, which asks about the type of abnormalities present in a certain region.}
    \label{fig:vqa-prompt}
\end{figure}

\section{Availability of Rectal MRI Dataset}
\label{app2}
\setcounter{figure}{0}
\setcounter{table}{0}

The in-house rectal MRI dataset consists of volumes acquired using three different contrasts: T1, T2 and DWI. For T1, T2 volumes, each volume was taken with one of Turbo Spin Echo (TSE), Fast Spin Echo (FSE) and Fast recovery FSE (FrFSE) imaging techniques. Only the T1, T2 volumes were used for training MS-VLM.

\begin{table}[H]
\centering
\caption{\\
Availability of in-house rectal MRI dataset.}
\label{tab:mri_imaging}
\resizebox{\textwidth}{!}{%
\begin{tabular}{lllcccc}
\toprule
\textbf{Contrast} & \textbf{Imaging Technique} & \textbf{Modifier} & \multicolumn{3}{c}{\textbf{Imaging Plane}} \\ 
\cmidrule(lr){4-6}
 & & & \textbf{Axial (+oblique)} & \textbf{Sagittal} & \textbf{Coronal (+oblique)} \\ \midrule
\multirow{2}{*}{\textbf{T1-weighted}} & TSE/FSE & None          & \cmark  & \cmark  & \xmark  \\
                                      & TSE/FSE & FS/SPIR, CE   & \cmark  & \cmark  & \cmark  \\ \midrule
\textbf{T2-weighted} & TSE/FSE/FrFSE & None    & \cmark  & \cmark  & \cmark  \\ \midrule
\multirow{3}{*}{\textbf{DWI}} & ADC         & N/A           & \cmark  & \xmark  & \xmark  \\
                              & b-value 0    & N/A           & \cmark  & \xmark  & \xmark  \\
                              & b-value (800 or 1000) & N/A   & \cmark  & \xmark  & \xmark  \\ \bottomrule
\end{tabular}%
}
\end{table}

\clearpage
\section{Instruction Prompts for LLM-based Evaluation}
\label{app3}
\setcounter{figure}{0}
\setcounter{table}{0}

\begin{figure}[H]
    \centering
    \includegraphics[width=1.0\textwidth, height=0.8\textheight, keepaspectratio]{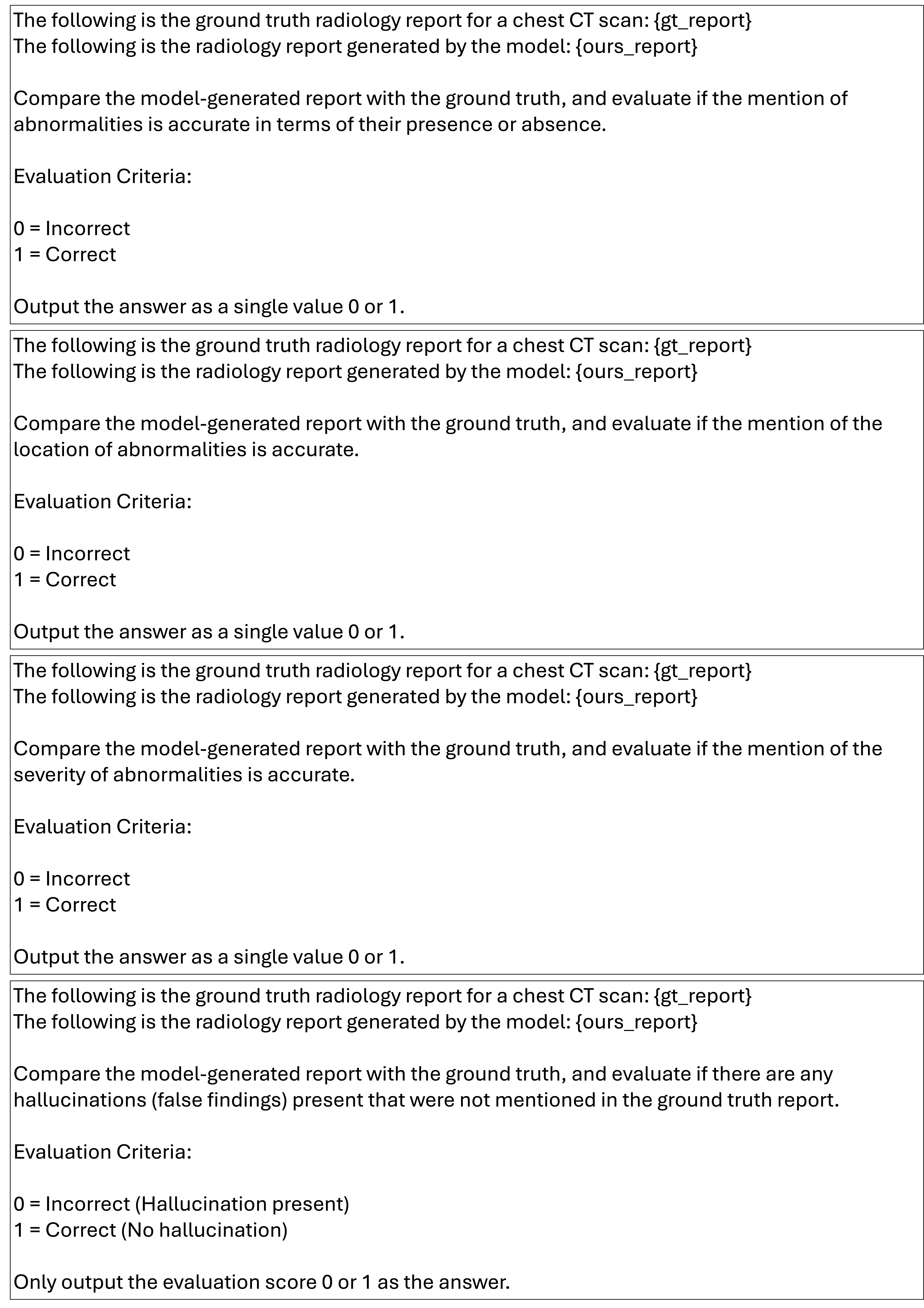}
    \caption{Instruction prompt for GPT-4o-mini used to evaluate the generated reports on the CT-RATE test set.}
    \label{fig:gpt4omini-prompt}
\end{figure}

\begin{figure}
    \centering
    \includegraphics[width=1.0\textwidth]{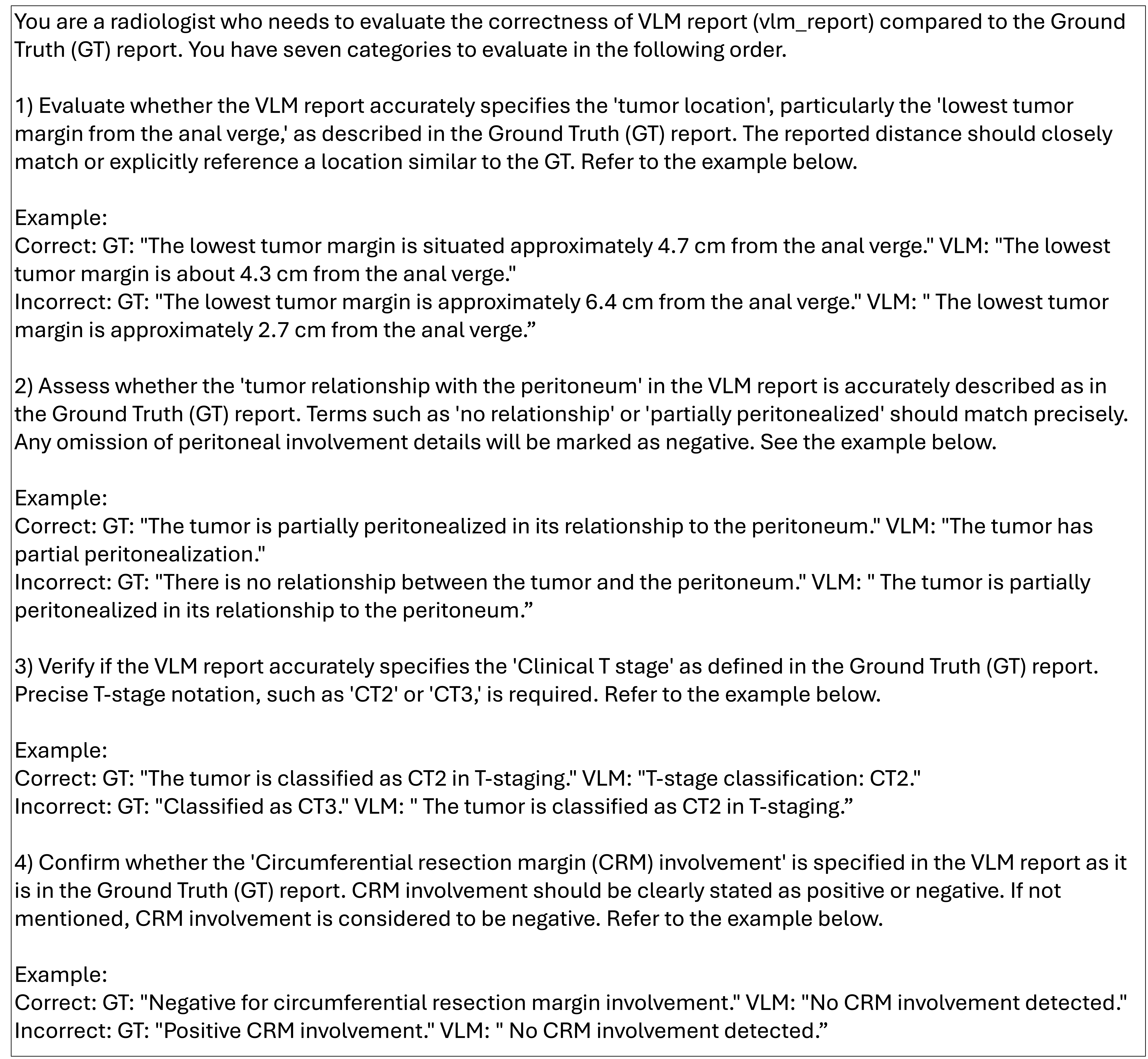}
\end{figure}

\begin{figure}
    \centering
    \includegraphics[width=1.0\textwidth]{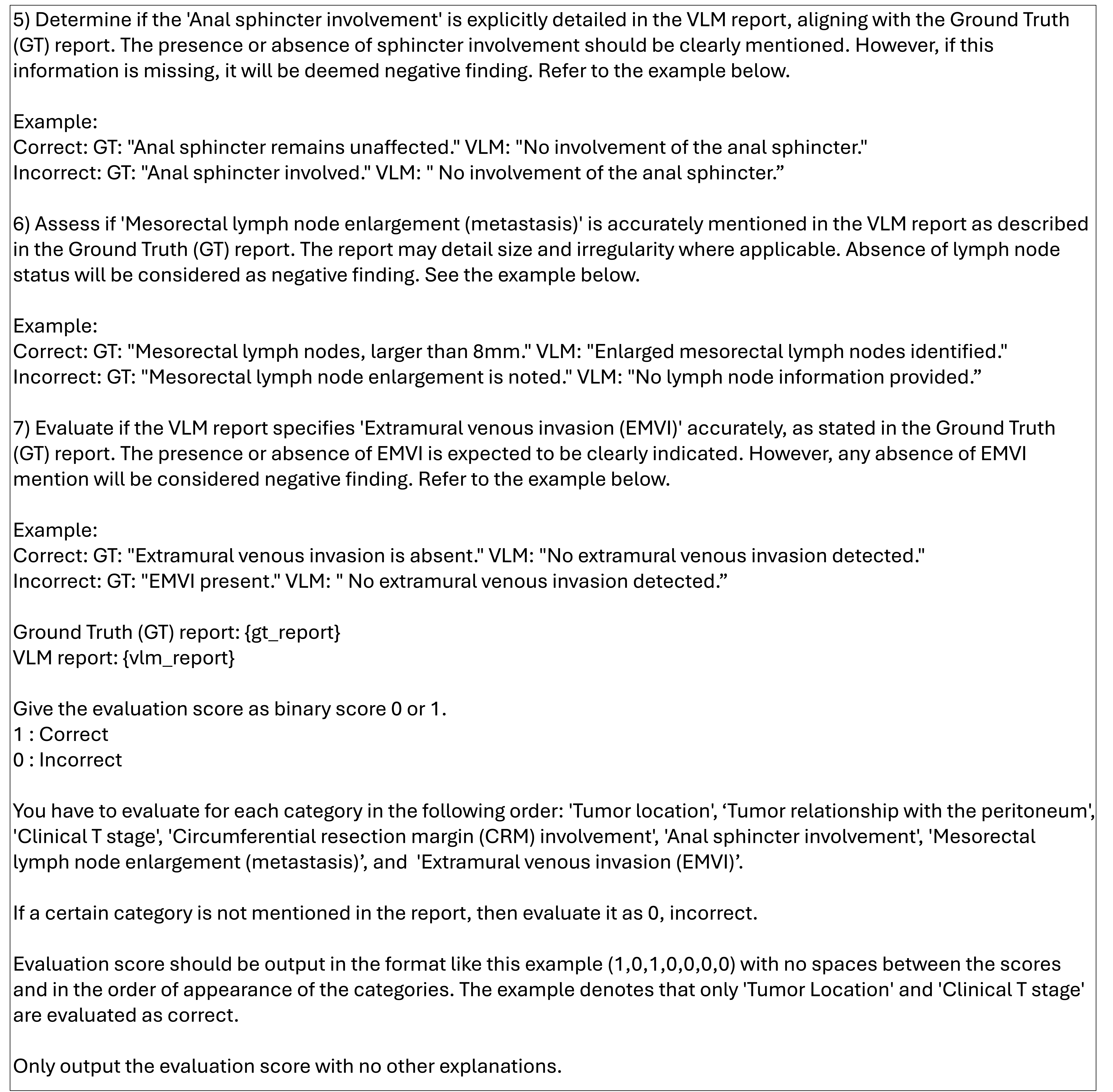}
    \caption{Instruction prompt for o1-mini used to evaluate the generated reports on the rectal MRI test set. The generated report is evaluated on seven abnormalities: Tumor location, Peritoneal involvement, T staging, CRM, ASI, MLNI, and EMVI.}
    \label{fig:o1mini-prompt2}
\end{figure}

\section{Qualitative Examples}
\label{app4}
\setcounter{figure}{0}
\setcounter{table}{0}

\begin{figure}[H]
    \includegraphics[width=1.0\textwidth, height=0.8\textheight, keepaspectratio]{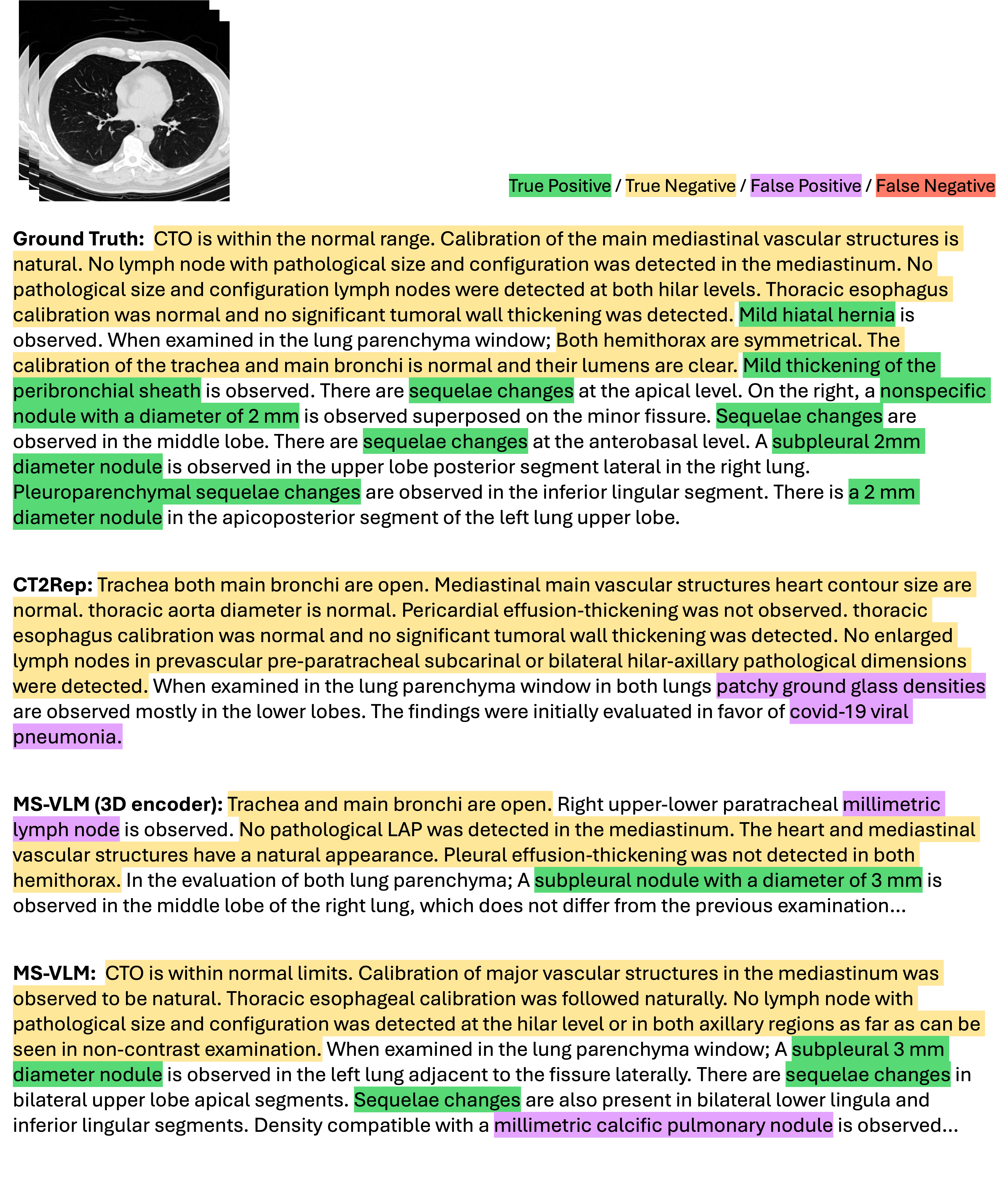}
    \centering
    \caption{Qualitative evaluation on CT-RATE. Each sentence from the report is color coded to identify True Positive, True Negative, False Positive and False Negatives on the findings mentioned in the ground truth report. MS-VLM achieves greater recall on the true positives.}
    \label{fig:qualitative-example2}
\end{figure}

\begin{figure}
    \includegraphics[width=1.0\textwidth, height=0.9\textheight, keepaspectratio]{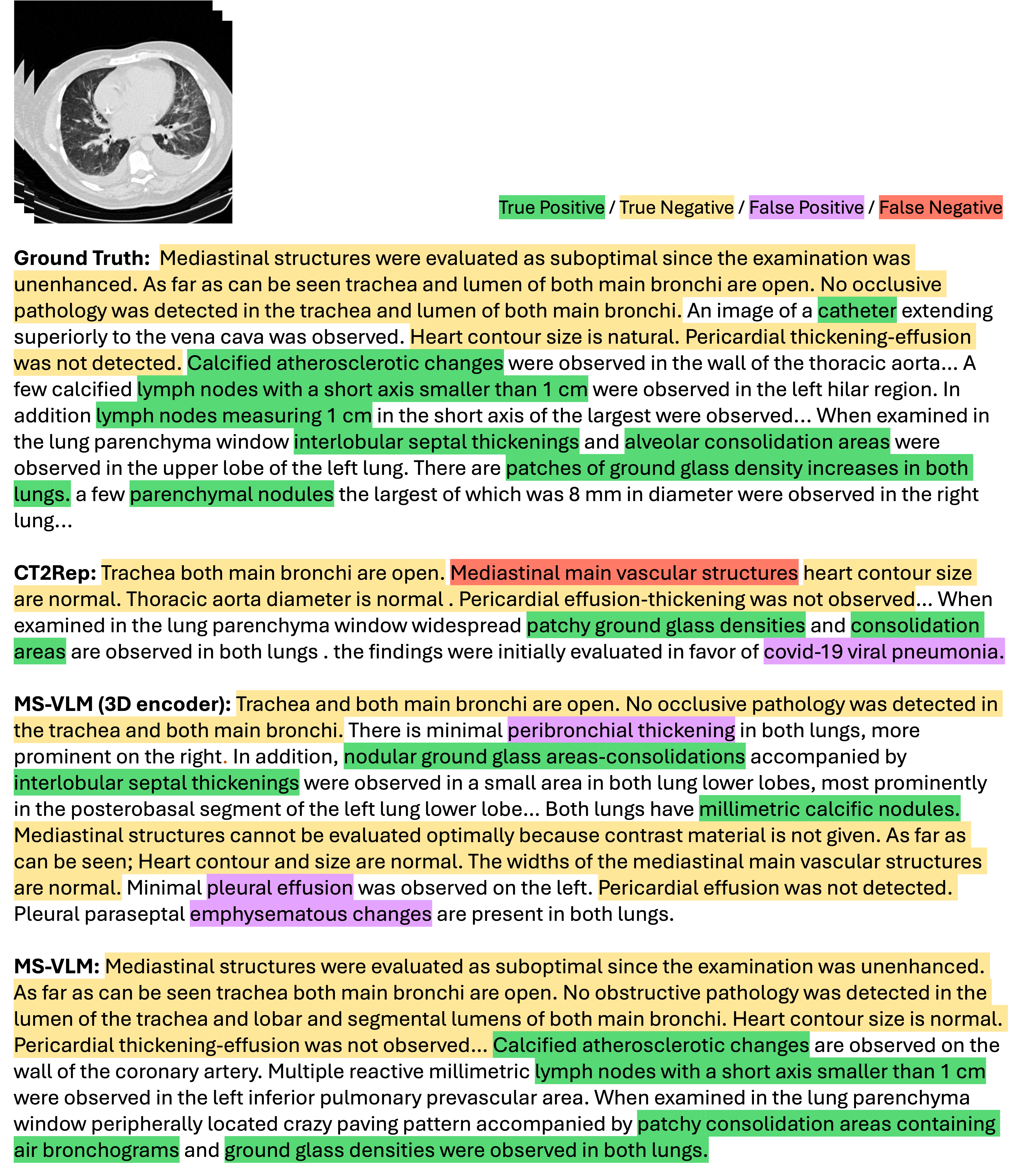}
    \centering
    \caption{Qualitative evaluation on CT-RATE. Each sentence from the report is color coded to identify True Positive, True Negative, False Positive and False Negatives on the findings mentioned in the ground truth report. MS-VLM achieves greater recall on the true positives while generating less false positive findings.}
    \label{fig:qualitative-example3}
\end{figure}

\begin{figure}
    \centering
    \includegraphics[width=0.8\textwidth]{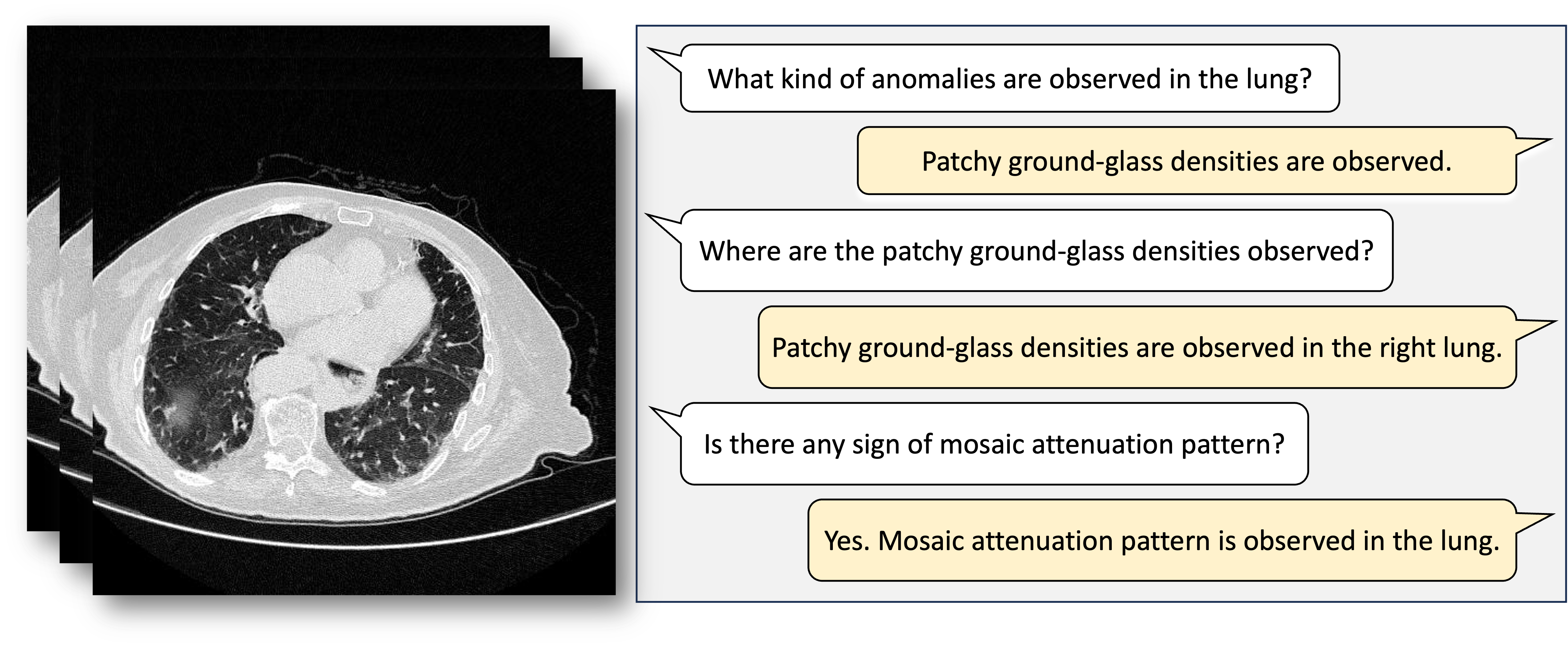}
    \caption{An example of VQA by MS-VLM on a test set sample of CT-RATE.}
    \label{fig:vqa_example}
\end{figure}

\begin{figure}
    \includegraphics[width=1.0\textwidth, height=1.0\textheight, keepaspectratio]{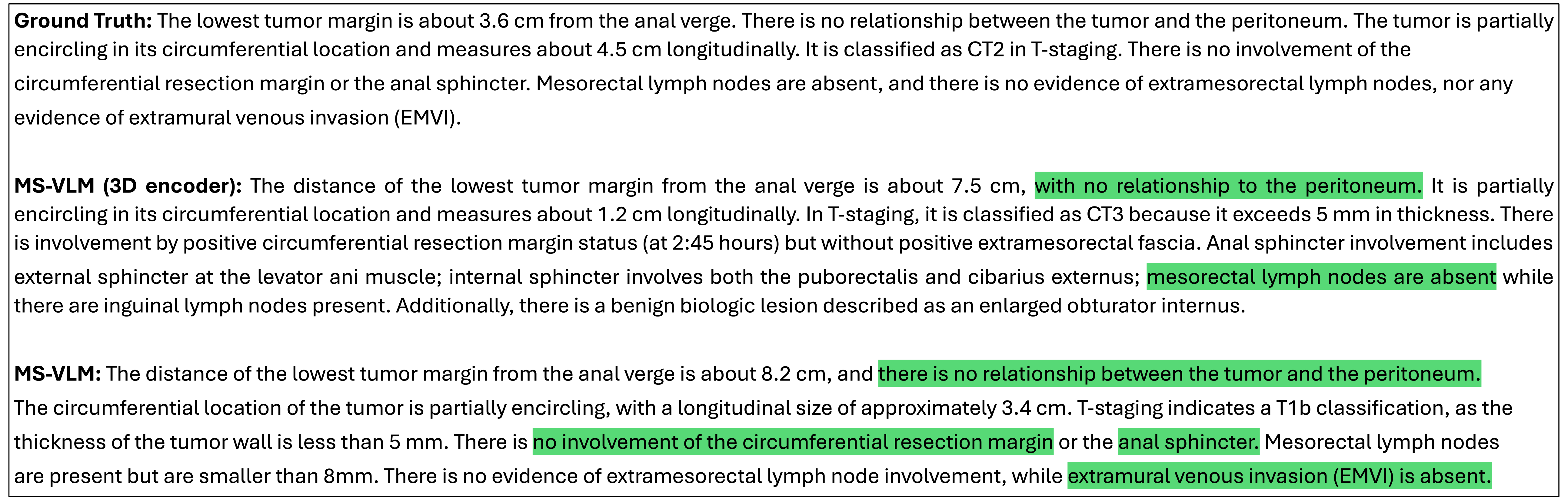}
    \centering
    \caption{Qualitative evaluation on rectal MRI. Correct predictions are highlighted. MS-VLM generates reports more faithful to the ground truth in terms of abnormality prediction.}
    \label{fig:qualitative-example4}
\end{figure}

\begin{figure}
    \includegraphics[width=1.0\textwidth, height=1.0\textheight, keepaspectratio]{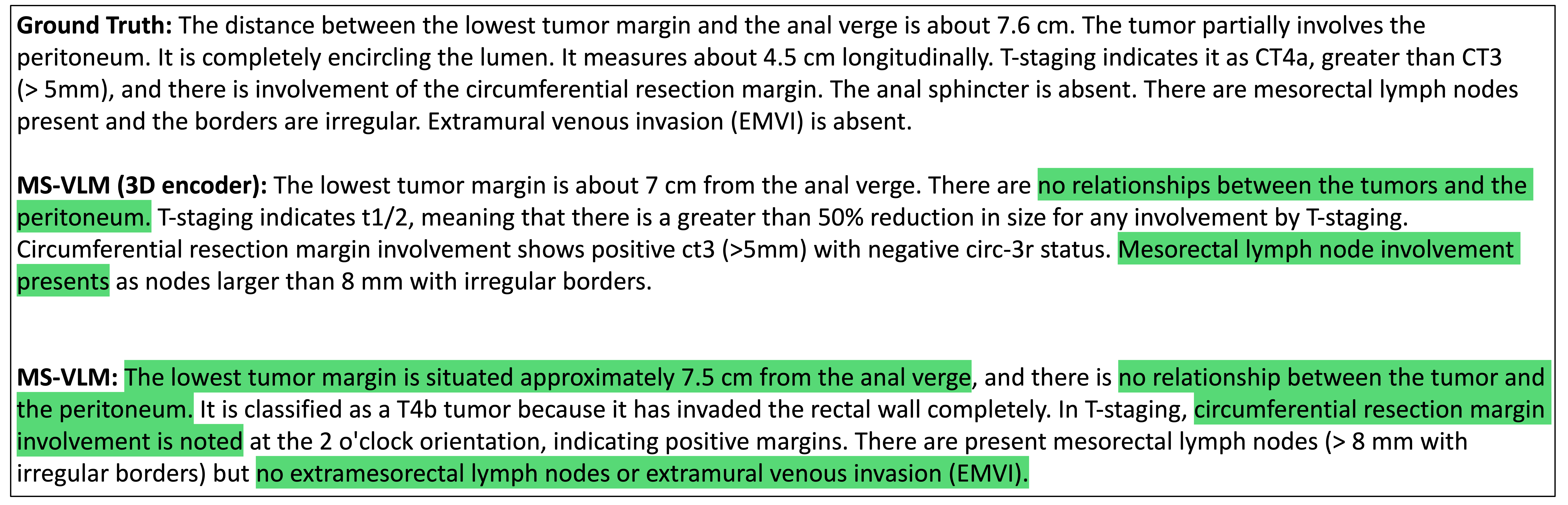}
    \centering
    \caption{Qualitative evaluation on rectal MRI. Correct predictions are highlighted. MS-VLM generates reports more faithful to the ground truth in terms of abnormality prediction.}
    \label{fig:qualitative-example5}
\end{figure}

\section{CA metrics on Chest CT Report Generation \& VQA}
\label{app5}
\setcounter{figure}{0}
\setcounter{table}{0}
    
\begin{table}[H]
    \centering
    \resizebox{0.75\textwidth}{!}{%
    \begin{threeparttable}
    \caption{\\ 
    Precision, Recall, and F1 Score across 18 chest-related abnormalities on the generated reports by MS-VLM.}
    \begin{tabular}{l c c c}
        \hline
        \textbf{Abnormality} & \textbf{Precision} & \textbf{Recall} & \textbf{F1 Score} \\ 
        \hline
        Medical material & 0.155 & 0.158 & 0.157 \\
        Arterial wall calcification & 0.306 & 0.367 & 0.334 \\
        Cardiomegaly & 0.249 & 0.298 & 0.271 \\
        Pericardial effusion & 0.130 & 0.214 & 0.162 \\
        Coronary artery wall calcification & 0.254 & 0.435 & 0.321 \\
        Hiatal hernia & 0.141 & 0.245 & 0.179 \\
        Lymphadenopathy & 0.268 & 0.393 & 0.319 \\
        Emphysema & 0.215 & 0.410 & 0.282 \\
        Atelectasis & 0.257 & 0.305 & 0.279 \\
        Lung nodule & 0.466 & 0.536 & 0.499 \\
        Lung opacity & 0.440 & 0.592 & 0.505 \\
        Pulmonary fibrotic sequela & 0.256 & 0.451 & 0.327 \\
        Pleural effusion & 0.133 & 0.315 & 0.187 \\
        Mosaic attenuation pattern & 0.128 & 0.230 & 0.164 \\
        Peribronchial thickening & 0.133 & 0.353 & 0.194 \\
        Consolidation & 0.235 & 0.370 & 0.287 \\
        Bronchiectasis & 0.125 & 0.100 & 0.111 \\
        Interlobular septal thickening & 0.104 & 0.157 & 0.125 \\
        \hline
        Average & 0.222 & 0.329 & 0.261 \\
        \hline
    \end{tabular}
    \end{threeparttable}}
    \label{tab:abnormality_metrics_report}
\end{table}

\begin{table}[H]
    \centering
    \resizebox{0.75\textwidth}{!}{%
    \begin{threeparttable}
    \caption{\\
    Precision, Recall, and F1 Score across 18 chest-related abnormalities on VQA by MS-VLM}
    \begin{tabular}{l c c c}
        \hline
        \textbf{Abnormality} & \textbf{Precision} & \textbf{Recall} & \textbf{F1 Score} \\
        \hline
        Medical material & 0.132 & 0.158 & 0.144 \\
        Arterial wall calcification & 0.546 & 0.436 & 0.485 \\
        Cardiomegaly & 0.143 & 0.198 & 0.166 \\
        Pericardial effusion & 0.674 & 0.602 & 0.636 \\
        Coronary artery wall calcification & 0.250 & 0.435 & 0.317 \\
        Hiatal hernia & 0.226 & 0.289 & 0.254 \\
        Lymphadenopathy & 0.293 & 0.537 & 0.379 \\
        Emphysema & 0.232 & 0.346 & 0.278 \\
        Atelectasis & 0.279 & 0.379 & 0.321 \\
        Lung nodule & 0.466 & 0.536 & 0.499 \\
        Lung opacity & 0.440 & 0.592 & 0.505 \\
        Pulmonary fibrotic sequela & 0.265 & 0.368 & 0.308 \\
        Pleural effusion & 0.200 & 0.287 & 0.236 \\
        Mosaic attenuation pattern & 0.105 & 0.185 & 0.134 \\
        Peribronchial thickening & 0.133 & 0.180 & 0.153 \\
        Consolidation & 0.252 & 0.292 & 0.271 \\
        Bronchiectasis & 0.178 & 0.096 & 0.125 \\
        Interlobular septal thickening & 0.132 & 0.156 & 0.143 \\
        \hline
        Average & 0.277 & 0.337 & 0.297 \\
        \hline
    \end{tabular}
    \end{threeparttable}}
    \label{tab:abnormality_metrics_vqa}
\end{table}

\bibliographystyle{elsarticle-harv} 
\bibliography{refs.bib}

\begin{thebibliography}{48}
\expandafter\ifx\csname natexlab\endcsname\relax\def\natexlab#1{#1}\fi
\providecommand{\url}[1]{\texttt{#1}}
\providecommand{\href}[2]{#2}
\providecommand{\path}[1]{#1}
\providecommand{\DOIprefix}{doi:}
\providecommand{\ArXivprefix}{arXiv:}
\providecommand{\URLprefix}{URL: }
\providecommand{\Pubmedprefix}{pmid:}
\providecommand{\doi}[1]{\href{http://dx.doi.org/#1}{\path{#1}}}
\providecommand{\Pubmed}[1]{\href{pmid:#1}{\path{#1}}}
\providecommand{\bibinfo}[2]{#2}
\ifx\xfnm\relax \def\xfnm[#1]{\unskip,\space#1}\fi
\bibitem[{Alayrac et~al.(2022)Alayrac, Donahue, Luc, Miech, Barr, Hasson, Lenc, Mensch, Millican, Reynolds et~al.}]{alayrac2022flamingo}
\bibinfo{author}{Alayrac, J.B.}, \bibinfo{author}{Donahue, J.}, \bibinfo{author}{Luc, P.}, \bibinfo{author}{Miech, A.}, \bibinfo{author}{Barr, I.}, \bibinfo{author}{Hasson, Y.}, \bibinfo{author}{Lenc, K.}, \bibinfo{author}{Mensch, A.}, \bibinfo{author}{Millican, K.}, \bibinfo{author}{Reynolds, M.}, et~al., \bibinfo{year}{2022}.
\newblock \bibinfo{title}{Flamingo: a visual language model for few-shot learning}.
\newblock \bibinfo{journal}{Advances in neural information processing systems} \bibinfo{volume}{35}, \bibinfo{pages}{23716--23736}.
\bibitem[{Bai et~al.(2024)Bai, Du, Huang, Meng and Zhao}]{bai2024m3d}
\bibinfo{author}{Bai, F.}, \bibinfo{author}{Du, Y.}, \bibinfo{author}{Huang, T.}, \bibinfo{author}{Meng, M.Q.H.}, \bibinfo{author}{Zhao, B.}, \bibinfo{year}{2024}.
\newblock \bibinfo{title}{M3d: Advancing 3d medical image analysis with multi-modal large language models}.
\newblock \bibinfo{journal}{arXiv preprint arXiv:2404.00578} .
\bibitem[{Blankemeier et~al.(2024)Blankemeier, Cohen, Kumar, Van~Veen, Gardezi, Paschali, Chen, Delbrouck, Reis, Truyts et~al.}]{blankemeier2024merlin}
\bibinfo{author}{Blankemeier, L.}, \bibinfo{author}{Cohen, J.P.}, \bibinfo{author}{Kumar, A.}, \bibinfo{author}{Van~Veen, D.}, \bibinfo{author}{Gardezi, S.J.S.}, \bibinfo{author}{Paschali, M.}, \bibinfo{author}{Chen, Z.}, \bibinfo{author}{Delbrouck, J.B.}, \bibinfo{author}{Reis, E.}, \bibinfo{author}{Truyts, C.}, et~al., \bibinfo{year}{2024}.
\newblock \bibinfo{title}{Merlin: A vision language foundation model for 3d computed tomography}.
\newblock \bibinfo{journal}{arXiv preprint arXiv:2406.06512} .
\bibitem[{Brady(2017)}]{brady2017error}
\bibinfo{author}{Brady, A.P.}, \bibinfo{year}{2017}.
\newblock \bibinfo{title}{Error and discrepancy in radiology: inevitable or avoidable?}
\newblock \bibinfo{journal}{Insights into Imaging} \bibinfo{volume}{8}, \bibinfo{pages}{171--182}.
\bibitem[{Bushberg and Boone(2011)}]{bushberg2011essential}
\bibinfo{author}{Bushberg, J.T.}, \bibinfo{author}{Boone, J.M.}, \bibinfo{year}{2011}.
\newblock \bibinfo{title}{The Essential Physics of Medical Imaging}.
\newblock \bibinfo{publisher}{Lippincott Williams \& Wilkins}.
\bibitem[{Caron et~al.(2021)Caron, Touvron, Misra, J{\'e}gou, Mairal, Bojanowski and Joulin}]{caron2021emerging}
\bibinfo{author}{Caron, M.}, \bibinfo{author}{Touvron, H.}, \bibinfo{author}{Misra, I.}, \bibinfo{author}{J{\'e}gou, H.}, \bibinfo{author}{Mairal, J.}, \bibinfo{author}{Bojanowski, P.}, \bibinfo{author}{Joulin, A.}, \bibinfo{year}{2021}.
\newblock \bibinfo{title}{Emerging properties in self-supervised vision transformers}, in: \bibinfo{booktitle}{Proceedings of the IEEE/CVF international conference on computer vision}, pp. \bibinfo{pages}{9650--9660}.
\bibitem[{Chaves et~al.(2024)Chaves, Huang, Xu, Xu, Usuyama, Zhang, Wang, Xie, Khademi, Yang et~al.}]{chaves2024training}
\bibinfo{author}{Chaves, J.M.Z.}, \bibinfo{author}{Huang, S.C.}, \bibinfo{author}{Xu, Y.}, \bibinfo{author}{Xu, H.}, \bibinfo{author}{Usuyama, N.}, \bibinfo{author}{Zhang, S.}, \bibinfo{author}{Wang, F.}, \bibinfo{author}{Xie, Y.}, \bibinfo{author}{Khademi, M.}, \bibinfo{author}{Yang, Z.}, et~al., \bibinfo{year}{2024}.
\newblock \bibinfo{title}{Training small multimodal models to bridge biomedical competency gap: A case study in radiology imaging}.
\newblock \bibinfo{journal}{CoRR} .
\bibitem[{Chen et~al.(2024)Chen, Zhao, Li, Zhong, Wang, Shang, Guo, Han, Liu, Liu et~al.}]{chen20243d}
\bibinfo{author}{Chen, H.}, \bibinfo{author}{Zhao, W.}, \bibinfo{author}{Li, Y.}, \bibinfo{author}{Zhong, T.}, \bibinfo{author}{Wang, Y.}, \bibinfo{author}{Shang, Y.}, \bibinfo{author}{Guo, L.}, \bibinfo{author}{Han, J.}, \bibinfo{author}{Liu, T.}, \bibinfo{author}{Liu, J.}, et~al., \bibinfo{year}{2024}.
\newblock \bibinfo{title}{3d-ct-gpt: Generating 3d radiology reports through integration of large vision-language models}.
\newblock \bibinfo{journal}{arXiv preprint arXiv:2409.19330} .
\bibitem[{Chiang et~al.(2023)Chiang, Li, Lin, Sheng, Wu, Zhang, Zheng, Zhuang, Zhuang, Gonzalez et~al.}]{chiang2023vicuna}
\bibinfo{author}{Chiang, W.L.}, \bibinfo{author}{Li, Z.}, \bibinfo{author}{Lin, Z.}, \bibinfo{author}{Sheng, Y.}, \bibinfo{author}{Wu, Z.}, \bibinfo{author}{Zhang, H.}, \bibinfo{author}{Zheng, L.}, \bibinfo{author}{Zhuang, S.}, \bibinfo{author}{Zhuang, Y.}, \bibinfo{author}{Gonzalez, J.E.}, et~al., \bibinfo{year}{2023}.
\newblock \bibinfo{title}{Vicuna: An open-source chatbot impressing gpt-4 with 90\%* chatgpt quality}.
\newblock \bibinfo{journal}{See https://vicuna. lmsys. org (accessed 14 April 2023)} \bibinfo{volume}{2}, \bibinfo{pages}{6}.
\bibitem[{Demner-Fushman et~al.(2012)Demner-Fushman, Antani, Simpson and Thoma}]{openi}
\bibinfo{author}{Demner-Fushman, D.}, \bibinfo{author}{Antani, S.}, \bibinfo{author}{Simpson, M.}, \bibinfo{author}{Thoma, G.}, \bibinfo{year}{2012}.
\newblock \bibinfo{title}{Design and development of a multimodal biomedical information retrieval system}.
\newblock \bibinfo{journal}{Journal of Computing Science and Engineering} \bibinfo{volume}{6}.
\newblock \DOIprefix\doi{10.5626/JCSE.2012.6.2.168}.
\bibitem[{Dosovitskiy(2020)}]{dosovitskiy2020image}
\bibinfo{author}{Dosovitskiy, A.}, \bibinfo{year}{2020}.
\newblock \bibinfo{title}{An image is worth 16x16 words: Transformers for image recognition at scale}.
\newblock \bibinfo{journal}{arXiv preprint arXiv:2010.11929} .
\bibitem[{Dou et~al.(2017)Dou, Yu, Chen, Jin, Yang, Qin and Heng}]{dou20173d}
\bibinfo{author}{Dou, Q.}, \bibinfo{author}{Yu, L.}, \bibinfo{author}{Chen, H.}, \bibinfo{author}{Jin, Y.}, \bibinfo{author}{Yang, X.}, \bibinfo{author}{Qin, J.}, \bibinfo{author}{Heng, P.A.}, \bibinfo{year}{2017}.
\newblock \bibinfo{title}{3d deeply supervised network for automated segmentation of volumetric medical images}.
\newblock \bibinfo{journal}{Medical image analysis} \bibinfo{volume}{41}, \bibinfo{pages}{40--54}.
\bibitem[{Hamamci et~al.(2024a)Hamamci, Er, Almas, Simsek, Esirgun, Dogan, Dasdelen, Wittmann, Simsar, Simsar et~al.}]{hamamci2024foundation}
\bibinfo{author}{Hamamci, I.E.}, \bibinfo{author}{Er, S.}, \bibinfo{author}{Almas, F.}, \bibinfo{author}{Simsek, A.G.}, \bibinfo{author}{Esirgun, S.N.}, \bibinfo{author}{Dogan, I.}, \bibinfo{author}{Dasdelen, M.F.}, \bibinfo{author}{Wittmann, B.}, \bibinfo{author}{Simsar, E.}, \bibinfo{author}{Simsar, M.}, et~al., \bibinfo{year}{2024}a.
\newblock \bibinfo{title}{A foundation model utilizing chest ct volumes and radiology reports for supervised-level zero-shot detection of abnormalities}.
\newblock \bibinfo{journal}{arXiv preprint arXiv:2403.17834} .
\bibitem[{Hamamci et~al.(2024b)Hamamci, Er and Menze}]{hamamci2024ct2rep}
\bibinfo{author}{Hamamci, I.E.}, \bibinfo{author}{Er, S.}, \bibinfo{author}{Menze, B.}, \bibinfo{year}{2024}b.
\newblock \bibinfo{title}{Ct2rep: Automated radiology report generation for 3d medical imaging}, in: \bibinfo{booktitle}{International Conference on Medical Image Computing and Computer-Assisted Intervention}, \bibinfo{organization}{Springer}. pp. \bibinfo{pages}{476--486}.
\bibitem[{Hatamizadeh et~al.(2022)Hatamizadeh, Tang, Nath, Yang, Myronenko, Landman, Roth and Xu}]{hatamizadeh2022unetr}
\bibinfo{author}{Hatamizadeh, A.}, \bibinfo{author}{Tang, Y.}, \bibinfo{author}{Nath, V.}, \bibinfo{author}{Yang, D.}, \bibinfo{author}{Myronenko, A.}, \bibinfo{author}{Landman, B.}, \bibinfo{author}{Roth, H.R.}, \bibinfo{author}{Xu, D.}, \bibinfo{year}{2022}.
\newblock \bibinfo{title}{Unetr: Transformers for 3d medical image segmentation}, in: \bibinfo{booktitle}{Proceedings of the IEEE/CVF winter conference on applications of computer vision}, pp. \bibinfo{pages}{574--584}.
\bibitem[{Hu et~al.(2024)Hu, Qian, Pan, Li, Qiu and Yang}]{hu2024advancing}
\bibinfo{author}{Hu, M.}, \bibinfo{author}{Qian, J.}, \bibinfo{author}{Pan, S.}, \bibinfo{author}{Li, Y.}, \bibinfo{author}{Qiu, R.L.}, \bibinfo{author}{Yang, X.}, \bibinfo{year}{2024}.
\newblock \bibinfo{title}{Advancing medical imaging with language models: featuring a spotlight on chatgpt}.
\newblock \bibinfo{journal}{Physics in Medicine \& Biology} \bibinfo{volume}{69}, \bibinfo{pages}{10TR01}.
\bibitem[{Irvin et~al.(2019)Irvin, Rajpurkar, Ko, Yu, Ciurea-Ilcus, Chute, Marklund, Haghgoo, Ball, Shpanskaya et~al.}]{irvin2019chexpert}
\bibinfo{author}{Irvin, J.}, \bibinfo{author}{Rajpurkar, P.}, \bibinfo{author}{Ko, M.}, \bibinfo{author}{Yu, Y.}, \bibinfo{author}{Ciurea-Ilcus, S.}, \bibinfo{author}{Chute, C.}, \bibinfo{author}{Marklund, H.}, \bibinfo{author}{Haghgoo, B.}, \bibinfo{author}{Ball, R.}, \bibinfo{author}{Shpanskaya, K.}, et~al., \bibinfo{year}{2019}.
\newblock \bibinfo{title}{Chexpert: A large chest radiograph dataset with uncertainty labels and expert comparison}, in: \bibinfo{booktitle}{Proceedings of the AAAI conference on artificial intelligence}, pp. \bibinfo{pages}{590--597}.
\bibitem[{Isensee et~al.(2021)Isensee, Jaeger, Kohl, Petersen and Maier-Hein}]{isensee2021nnu}
\bibinfo{author}{Isensee, F.}, \bibinfo{author}{Jaeger, P.F.}, \bibinfo{author}{Kohl, S.A.}, \bibinfo{author}{Petersen, J.}, \bibinfo{author}{Maier-Hein, K.H.}, \bibinfo{year}{2021}.
\newblock \bibinfo{title}{nnu-net: a self-configuring method for deep learning-based biomedical image segmentation}.
\newblock \bibinfo{journal}{Nature methods} \bibinfo{volume}{18}, \bibinfo{pages}{203--211}.
\bibitem[{Johnson et~al.(2019)Johnson, Pollard, Berkowitz, Greenbaum, Lungren, Deng, Mark and Horng}]{mimiccxr}
\bibinfo{author}{Johnson, A.}, \bibinfo{author}{Pollard, T.}, \bibinfo{author}{Berkowitz, S.}, \bibinfo{author}{Greenbaum, N.}, \bibinfo{author}{Lungren, M.}, \bibinfo{author}{Deng, C.y.}, \bibinfo{author}{Mark, R.}, \bibinfo{author}{Horng, S.}, \bibinfo{year}{2019}.
\newblock \bibinfo{title}{Mimic-cxr, a de-identified publicly available database of chest radiographs with free-text reports}.
\newblock \bibinfo{journal}{Scientific Data} \bibinfo{volume}{6}, \bibinfo{pages}{317}.
\newblock \DOIprefix\doi{10.1038/s41597-019-0322-0}.
\bibitem[{Kalender(2011)}]{kalender2011computed}
\bibinfo{author}{Kalender, W.A.}, \bibinfo{year}{2011}.
\newblock \bibinfo{title}{Computed Tomography: Fundamentals, System Technology, Image Quality, Applications}.
\newblock \bibinfo{publisher}{Publicis Publishing}.
\bibitem[{Lavie and Denkowski(2009)}]{lavie2009meteor}
\bibinfo{author}{Lavie, A.}, \bibinfo{author}{Denkowski, M.J.}, \bibinfo{year}{2009}.
\newblock \bibinfo{title}{The meteor metric for automatic evaluation of machine translation}.
\newblock \bibinfo{journal}{Machine translation} \bibinfo{volume}{23}, \bibinfo{pages}{105--115}.
\bibitem[{Lee et~al.(2024)Lee, Kim, Chang and Ye}]{lee2024llmcxr}
\bibinfo{author}{Lee, S.}, \bibinfo{author}{Kim, W.J.}, \bibinfo{author}{Chang, J.}, \bibinfo{author}{Ye, J.C.}, \bibinfo{year}{2024}.
\newblock \bibinfo{title}{{LLM}-{CXR}: Instruction-finetuned {LLM} for {CXR} image understanding and generation}, in: \bibinfo{booktitle}{The Twelfth International Conference on Learning Representations}.
\newblock \URLprefix \url{https://openreview.net/forum?id=BqHaLnans2}.
\bibitem[{Li et~al.(2024)Li, Wong, Zhang, Usuyama, Liu, Yang, Naumann, Poon and Gao}]{li2024llava}
\bibinfo{author}{Li, C.}, \bibinfo{author}{Wong, C.}, \bibinfo{author}{Zhang, S.}, \bibinfo{author}{Usuyama, N.}, \bibinfo{author}{Liu, H.}, \bibinfo{author}{Yang, J.}, \bibinfo{author}{Naumann, T.}, \bibinfo{author}{Poon, H.}, \bibinfo{author}{Gao, J.}, \bibinfo{year}{2024}.
\newblock \bibinfo{title}{Llava-med: Training a large language-and-vision assistant for biomedicine in one day}.
\newblock \bibinfo{journal}{Advances in Neural Information Processing Systems} \bibinfo{volume}{36}.
\bibitem[{Litjens et~al.(2017)}]{litjens2017survey}
\bibinfo{author}{Litjens, G.}, et~al., \bibinfo{year}{2017}.
\newblock \bibinfo{title}{A survey on deep learning in medical image analysis}.
\newblock \bibinfo{journal}{Medical Image Analysis} \bibinfo{volume}{42}, \bibinfo{pages}{60--88}.
\bibitem[{Liu et~al.(2021)Liu, Wu, Ge, Fan and Zou}]{liu2021exploring}
\bibinfo{author}{Liu, F.}, \bibinfo{author}{Wu, X.}, \bibinfo{author}{Ge, S.}, \bibinfo{author}{Fan, W.}, \bibinfo{author}{Zou, Y.}, \bibinfo{year}{2021}.
\newblock \bibinfo{title}{Exploring and distilling posterior and prior knowledge for radiology report generation}, in: \bibinfo{booktitle}{Proceedings of the IEEE/CVF conference on computer vision and pattern recognition}, pp. \bibinfo{pages}{13753--13762}.
\bibitem[{Liu et~al.(2024)Liu, Li, Wu and Lee}]{liu2024visual}
\bibinfo{author}{Liu, H.}, \bibinfo{author}{Li, C.}, \bibinfo{author}{Wu, Q.}, \bibinfo{author}{Lee, Y.J.}, \bibinfo{year}{2024}.
\newblock \bibinfo{title}{Visual instruction tuning}.
\newblock \bibinfo{journal}{Advances in neural information processing systems} \bibinfo{volume}{36}.
\bibitem[{Lu et~al.(2024)Lu, Chen, Williamson, Chen, Zhao, Chow, Ikemura, Kim, Pouli, Patel et~al.}]{lu2024multimodal}
\bibinfo{author}{Lu, M.Y.}, \bibinfo{author}{Chen, B.}, \bibinfo{author}{Williamson, D.F.}, \bibinfo{author}{Chen, R.J.}, \bibinfo{author}{Zhao, M.}, \bibinfo{author}{Chow, A.K.}, \bibinfo{author}{Ikemura, K.}, \bibinfo{author}{Kim, A.}, \bibinfo{author}{Pouli, D.}, \bibinfo{author}{Patel, A.}, et~al., \bibinfo{year}{2024}.
\newblock \bibinfo{title}{A multimodal generative ai copilot for human pathology}.
\newblock \bibinfo{journal}{Nature} \bibinfo{volume}{634}, \bibinfo{pages}{466--473}.
\bibitem[{Madan et~al.(2024)Madan, M{\o}gelmose, Modi, Rawat and Moeslund}]{madan2024foundation}
\bibinfo{author}{Madan, N.}, \bibinfo{author}{M{\o}gelmose, A.}, \bibinfo{author}{Modi, R.}, \bibinfo{author}{Rawat, Y.S.}, \bibinfo{author}{Moeslund, T.B.}, \bibinfo{year}{2024}.
\newblock \bibinfo{title}{Foundation models for video understanding: A survey}.
\newblock \bibinfo{journal}{arXiv preprint arXiv:2405.03770} .
\bibitem[{McRobbie et~al.(2017)}]{mcrobbie2017mri}
\bibinfo{author}{McRobbie, D.W.}, et~al., \bibinfo{year}{2017}.
\newblock \bibinfo{title}{MRI from Picture to Proton}.
\newblock \bibinfo{publisher}{Cambridge University Press}.
\bibitem[{Moon et~al.(2022)Moon, Lee, Shin, Kim and Choi}]{moon2022multi}
\bibinfo{author}{Moon, J.H.}, \bibinfo{author}{Lee, H.}, \bibinfo{author}{Shin, W.}, \bibinfo{author}{Kim, Y.H.}, \bibinfo{author}{Choi, E.}, \bibinfo{year}{2022}.
\newblock \bibinfo{title}{Multi-modal understanding and generation for medical images and text via vision-language pre-training}.
\newblock \bibinfo{journal}{IEEE Journal of Biomedical and Health Informatics} \bibinfo{volume}{26}, \bibinfo{pages}{6070--6080}.
\bibitem[{Papineni et~al.(2002)Papineni, Roukos, Ward and Zhu}]{papineni2002bleu}
\bibinfo{author}{Papineni, K.}, \bibinfo{author}{Roukos, S.}, \bibinfo{author}{Ward, T.}, \bibinfo{author}{Zhu, W.J.}, \bibinfo{year}{2002}.
\newblock \bibinfo{title}{Bleu: a method for automatic evaluation of machine translation}, in: \bibinfo{booktitle}{Proceedings of the 40th annual meeting of the Association for Computational Linguistics}, pp. \bibinfo{pages}{311--318}.
\bibitem[{Park et~al.(2024)Park, Lee, Shin, Lee and Ye}]{park2024self}
\bibinfo{author}{Park, S.}, \bibinfo{author}{Lee, E.S.}, \bibinfo{author}{Shin, K.S.}, \bibinfo{author}{Lee, J.E.}, \bibinfo{author}{Ye, J.C.}, \bibinfo{year}{2024}.
\newblock \bibinfo{title}{Self-supervised multi-modal training from uncurated images and reports enables monitoring ai in radiology}.
\newblock \bibinfo{journal}{Medical Image Analysis} \bibinfo{volume}{91}, \bibinfo{pages}{103021}.
\bibitem[{Pham et~al.(2022)Pham, Nguyen, Nguyen, Le and Khanh}]{pham2022accurate}
\bibinfo{author}{Pham, H.H.}, \bibinfo{author}{Nguyen, H.Q.}, \bibinfo{author}{Nguyen, H.T.}, \bibinfo{author}{Le, L.T.}, \bibinfo{author}{Khanh, L.}, \bibinfo{year}{2022}.
\newblock \bibinfo{title}{An accurate and explainable deep learning system improves interobserver agreement in the interpretation of chest radiograph}.
\newblock \bibinfo{journal}{IEEE Access} \bibinfo{volume}{10}, \bibinfo{pages}{104512--104531}.
\bibitem[{Qu et~al.(2021)Qu, Zhou, Yan, Wang, Rustgi, Zhang, Gevaert and Metaxas}]{qu2021genetic}
\bibinfo{author}{Qu, H.}, \bibinfo{author}{Zhou, M.}, \bibinfo{author}{Yan, Z.}, \bibinfo{author}{Wang, H.}, \bibinfo{author}{Rustgi, V.K.}, \bibinfo{author}{Zhang, S.}, \bibinfo{author}{Gevaert, O.}, \bibinfo{author}{Metaxas, D.N.}, \bibinfo{year}{2021}.
\newblock \bibinfo{title}{Genetic mutation and biological pathway prediction based on whole slide images in breast carcinoma using deep learning}.
\newblock \bibinfo{journal}{NPJ precision oncology} \bibinfo{volume}{5}, \bibinfo{pages}{87}.
\bibitem[{Radford et~al.(2021)Radford, Kim, Hallacy, Ramesh, Goh, Agarwal, Sastry, Askell, Mishkin, Clark et~al.}]{radford2021learning}
\bibinfo{author}{Radford, A.}, \bibinfo{author}{Kim, J.W.}, \bibinfo{author}{Hallacy, C.}, \bibinfo{author}{Ramesh, A.}, \bibinfo{author}{Goh, G.}, \bibinfo{author}{Agarwal, S.}, \bibinfo{author}{Sastry, G.}, \bibinfo{author}{Askell, A.}, \bibinfo{author}{Mishkin, P.}, \bibinfo{author}{Clark, J.}, et~al., \bibinfo{year}{2021}.
\newblock \bibinfo{title}{Learning transferable visual models from natural language supervision}, in: \bibinfo{booktitle}{International conference on machine learning}, \bibinfo{organization}{PMLR}. pp. \bibinfo{pages}{8748--8763}.
\bibitem[{Rouge(2004)}]{rouge2004package}
\bibinfo{author}{Rouge, L.C.}, \bibinfo{year}{2004}.
\newblock \bibinfo{title}{A package for automatic evaluation of summaries}, in: \bibinfo{booktitle}{Proceedings of Workshop on Text Summarization of ACL, Spain}.
\bibitem[{Taori et~al.(2023)Taori, Gulrajani, Zhang, Dubois, Li, Guestrin, Liang and Hashimoto}]{taori2023stanford}
\bibinfo{author}{Taori, R.}, \bibinfo{author}{Gulrajani, I.}, \bibinfo{author}{Zhang, T.}, \bibinfo{author}{Dubois, Y.}, \bibinfo{author}{Li, X.}, \bibinfo{author}{Guestrin, C.}, \bibinfo{author}{Liang, P.}, \bibinfo{author}{Hashimoto, T.B.}, \bibinfo{year}{2023}.
\newblock \bibinfo{title}{Stanford alpaca: an instruction-following llama model (2023)}.
\newblock \bibinfo{journal}{URL https://github. com/tatsu-lab/stanford\_alpaca} \bibinfo{volume}{1}.
\bibitem[{Tiu et~al.(2022)Tiu, Talius, Patel, Langlotz, Ng and Rajpurkar}]{tiu2022expert}
\bibinfo{author}{Tiu, E.}, \bibinfo{author}{Talius, E.}, \bibinfo{author}{Patel, P.}, \bibinfo{author}{Langlotz, C.P.}, \bibinfo{author}{Ng, A.Y.}, \bibinfo{author}{Rajpurkar, P.}, \bibinfo{year}{2022}.
\newblock \bibinfo{title}{Expert-level detection of pathologies from unannotated chest x-ray images via self-supervised learning}.
\newblock \bibinfo{journal}{Nature Biomedical Engineering} \bibinfo{volume}{6}, \bibinfo{pages}{1399--1406}.
\bibitem[{Tu et~al.(2024)Tu, Azizi, Driess, Schaekermann, Amin, Chang, Carroll, Lau, Tanno, Ktena et~al.}]{tu2024towards}
\bibinfo{author}{Tu, T.}, \bibinfo{author}{Azizi, S.}, \bibinfo{author}{Driess, D.}, \bibinfo{author}{Schaekermann, M.}, \bibinfo{author}{Amin, M.}, \bibinfo{author}{Chang, P.C.}, \bibinfo{author}{Carroll, A.}, \bibinfo{author}{Lau, C.}, \bibinfo{author}{Tanno, R.}, \bibinfo{author}{Ktena, I.}, et~al., \bibinfo{year}{2024}.
\newblock \bibinfo{title}{Towards generalist biomedical ai}.
\newblock \bibinfo{journal}{NEJM AI} \bibinfo{volume}{1}, \bibinfo{pages}{AIoa2300138}.
\bibitem[{Wang et~al.(2019)Wang, Han, Chen, Gao and Vasconcelos}]{wang2019volumetric}
\bibinfo{author}{Wang, X.}, \bibinfo{author}{Han, S.}, \bibinfo{author}{Chen, Y.}, \bibinfo{author}{Gao, D.}, \bibinfo{author}{Vasconcelos, N.}, \bibinfo{year}{2019}.
\newblock \bibinfo{title}{Volumetric attention for 3d medical image segmentation and detection}, in: \bibinfo{booktitle}{Medical Image Computing and Computer Assisted Intervention--MICCAI 2019: 22nd International Conference, Shenzhen, China, October 13--17, 2019, Proceedings, Part VI 22}, \bibinfo{organization}{Springer}. pp. \bibinfo{pages}{175--184}.
\bibitem[{Wang et~al.(2022)Wang, Wu, Agarwal and Sun}]{wang2022medclip}
\bibinfo{author}{Wang, Z.}, \bibinfo{author}{Wu, Z.}, \bibinfo{author}{Agarwal, D.}, \bibinfo{author}{Sun, J.}, \bibinfo{year}{2022}.
\newblock \bibinfo{title}{Medclip: Contrastive learning from unpaired medical images and text}.
\newblock \bibinfo{journal}{arXiv preprint arXiv:2210.10163} .
\bibitem[{Wu et~al.(2023)Wu, Zhang, Zhang, Wang and Xie}]{wu2023towards}
\bibinfo{author}{Wu, C.}, \bibinfo{author}{Zhang, X.}, \bibinfo{author}{Zhang, Y.}, \bibinfo{author}{Wang, Y.}, \bibinfo{author}{Xie, W.}, \bibinfo{year}{2023}.
\newblock \bibinfo{title}{Towards generalist foundation model for radiology}.
\newblock \bibinfo{journal}{arXiv preprint arXiv:2308.02463} .
\bibitem[{Xie et~al.(2022)Xie, Zhang, Cao, Lin, Bao, Yao, Dai and Hu}]{xie2022simmim}
\bibinfo{author}{Xie, Z.}, \bibinfo{author}{Zhang, Z.}, \bibinfo{author}{Cao, Y.}, \bibinfo{author}{Lin, Y.}, \bibinfo{author}{Bao, J.}, \bibinfo{author}{Yao, Z.}, \bibinfo{author}{Dai, Q.}, \bibinfo{author}{Hu, H.}, \bibinfo{year}{2022}.
\newblock \bibinfo{title}{Simmim: A simple framework for masked image modeling}, in: \bibinfo{booktitle}{Proceedings of the IEEE/CVF conference on computer vision and pattern recognition}, pp. \bibinfo{pages}{9653--9663}.
\bibitem[{Yan et~al.(2022)Yan, McAuley, Lu, Du, Chang, Gentili and Hsu}]{yan2022radbert}
\bibinfo{author}{Yan, A.}, \bibinfo{author}{McAuley, J.}, \bibinfo{author}{Lu, X.}, \bibinfo{author}{Du, J.}, \bibinfo{author}{Chang, E.Y.}, \bibinfo{author}{Gentili, A.}, \bibinfo{author}{Hsu, C.N.}, \bibinfo{year}{2022}.
\newblock \bibinfo{title}{Radbert: adapting transformer-based language models to radiology}.
\newblock \bibinfo{journal}{Radiology: Artificial Intelligence} \bibinfo{volume}{4}, \bibinfo{pages}{e210258}.
\bibitem[{You et~al.(2023)You, Gu, Ham, Park, Kim, Hong, Baek and Roh}]{you2023cxr}
\bibinfo{author}{You, K.}, \bibinfo{author}{Gu, J.}, \bibinfo{author}{Ham, J.}, \bibinfo{author}{Park, B.}, \bibinfo{author}{Kim, J.}, \bibinfo{author}{Hong, E.K.}, \bibinfo{author}{Baek, W.}, \bibinfo{author}{Roh, B.}, \bibinfo{year}{2023}.
\newblock \bibinfo{title}{Cxr-clip: Toward large scale chest x-ray language-image pre-training}, in: \bibinfo{booktitle}{International Conference on Medical Image Computing and Computer-Assisted Intervention}, \bibinfo{organization}{Springer}. pp. \bibinfo{pages}{101--111}.
\bibitem[{Zaheer et~al.(2020)Zaheer, Guruganesh, Dubey, Ainslie, Alberti, Ontanon, Pham, Ravula, Wang, Yang et~al.}]{zaheer2020big}
\bibinfo{author}{Zaheer, M.}, \bibinfo{author}{Guruganesh, G.}, \bibinfo{author}{Dubey, K.A.}, \bibinfo{author}{Ainslie, J.}, \bibinfo{author}{Alberti, C.}, \bibinfo{author}{Ontanon, S.}, \bibinfo{author}{Pham, P.}, \bibinfo{author}{Ravula, A.}, \bibinfo{author}{Wang, Q.}, \bibinfo{author}{Yang, L.}, et~al., \bibinfo{year}{2020}.
\newblock \bibinfo{title}{Big bird: Transformers for longer sequences}.
\newblock \bibinfo{journal}{Advances in neural information processing systems} \bibinfo{volume}{33}, \bibinfo{pages}{17283--17297}.
\bibitem[{Zhang et~al.(2024a)Zhang, Zhou, Adhikarla, Yan, Liu, Yu, Liu, Chen, Davison, Ren et~al.}]{zhang2024generalist}
\bibinfo{author}{Zhang, K.}, \bibinfo{author}{Zhou, R.}, \bibinfo{author}{Adhikarla, E.}, \bibinfo{author}{Yan, Z.}, \bibinfo{author}{Liu, Y.}, \bibinfo{author}{Yu, J.}, \bibinfo{author}{Liu, Z.}, \bibinfo{author}{Chen, X.}, \bibinfo{author}{Davison, B.D.}, \bibinfo{author}{Ren, H.}, et~al., \bibinfo{year}{2024}a.
\newblock \bibinfo{title}{A generalist vision--language foundation model for diverse biomedical tasks}.
\newblock \bibinfo{journal}{Nature Medicine} , \bibinfo{pages}{1--13}.
\bibitem[{Zhang et~al.(2024b)Zhang, Wu, Zhao, Lei, Zhang, Wang and Xie}]{zhang2024radgenome}
\bibinfo{author}{Zhang, X.}, \bibinfo{author}{Wu, C.}, \bibinfo{author}{Zhao, Z.}, \bibinfo{author}{Lei, J.}, \bibinfo{author}{Zhang, Y.}, \bibinfo{author}{Wang, Y.}, \bibinfo{author}{Xie, W.}, \bibinfo{year}{2024}b.
\newblock \bibinfo{title}{Radgenome-chest ct: A grounded vision-language dataset for chest ct analysis}.
\newblock \bibinfo{journal}{arXiv preprint arXiv:2404.16754} .

\end{thebibliography}

\end{document}